\documentclass[onecolumn,12pt,journal,final]{IEEEtran}

\usepackage{amsfonts}
\usepackage[dvips]{graphicx}
\usepackage{times}
\usepackage{cite}
\usepackage{amsmath}
\usepackage{cases}
\usepackage{color}
\usepackage{array}
\usepackage{algorithm}
\usepackage{algorithmic}
\usepackage{dsfont}
\usepackage{amssymb}
\usepackage{stfloats}
\usepackage{graphicx}
\usepackage{footnote}
\usepackage{booktabs}
\usepackage{array}
\usepackage{bm}
\usepackage{stfloats}
\usepackage{diagbox}
\usepackage{multirow}

\newtheorem{theorem}{Theorem}
\newtheorem{remark}{Remark}

\newtheorem{lemma}{Lemma}

\begin{document}
\title{Degrees of Freedom of Cache-Aided Wireless Cellular Networks}
\author{\IEEEauthorblockN{Youlong Cao, \IEEEmembership{Student Member,~IEEE}, and Meixia Tao, \IEEEmembership{Fellow,~IEEE}\\}

\thanks{This work is supported by the NSF of China under grant 61941106 and the National Key R$\&$D Project of China under grant 2019YFB1802702. This work was presented in part at the 2019 IEEE Wireless Communications and Networking Conference \cite{long_wcnc19}. The authors are with Shanghai Institute for Advanced Communication and Data Science, and Department of Electrical Engineering, Shanghai Jiao Tong University, Shanghai 200240, China. Emails: \{caoyoulong, mxtao\}@sjtu.edu.cn.}
}
\maketitle

\begin{abstract}
This work investigates the degrees of freedom (DoF) of a downlink cache-aided cellular network where the locations of base stations (BSs) are modeled as a grid topology and users within a grid cell can only communicate with four nearby BSs. We adopt a cache placement method with uncoded prefetching tailored for the network with partial connectivity. According to the overlapped degree of cached contents among BSs, we propose transmission schemes with no BS cooperation, partial BS cooperation, and full BS cooperation, respectively, for different cache sizes. In specific, the common cached contents among BSs are utilized to cancel some undesired signals by interference neutralization while interference alignment is used to coordinate signals of distinct cached contents. Our achievable results indicate that the reciprocal of per-user DoF of the cellular network decreases piecewise linearly with the normalized cache size $\mu$ at each BS, and the gain of BS caching is more significant for the small cache region. Under the given cache placement scheme, we also provide an upper bound of per-user DoF and show that our achievable DoF is optimal when $\mu\in\left[\frac{1}{2},1\right]$, and within an additive gap of $\frac{4}{39}$ to the optimum when $\mu\in\left[\frac{1}{4},\frac{1}{2}\right)$.
\end{abstract}


\section{Introduction}
Content delivery applications such as video streaming have occupied a significant proportion in mobile wireless data traffic. It amounts for more than 63$\%$ of the total mobile data in 2019 and is foreseen to contribute 76$\%$ in 2025 \cite{ericsson}. Wireless caching \cite{caching1,caching2,liu_tao} is considered as one of the most effective techniques to cope with this increasing content oriented traffic. Its main idea is to exploit the under-utilized network resources during the off-peak hours by prefetching the popular contents in the local memory of edge nodes in order to accelerate the content delivery. Caching is first studied from an information-theoretic framework, known as coded caching \cite{Alilimits} where a server communicates with multiple cache-enabled receivers over a shared error-free link. Receiver caching has been shown as an effective way to reduce the traffic load by creating coded multicast opportunities for the memory-equipped networks \cite{Alilimits,Soneze_distributed_computing,Kiamari_distributed_computing}. Then the authors in \cite{Ali2} investigate the role of transmitter caching in a $3\times 3$ interference channel, and show that transmitter caching can bring transmit cooperation gain and hence provide an opportunity to increase the degrees of freedom (DoF) of the interference channel. The works \cite{xu,Ali3,niesendof} consider caching in a general cache-aided interference network with arbitrary number of transmitters and arbitrary number of receivers. It is found in \cite{xu} that with a novel cooperative transmitter and receiver caching strategy, the interference network can be turned opportunistically into a new class of channels, namely, cooperative X-multicast channels. The works \cite{long_journal,long_journal_gms,MISO_Petros_Elia,yangsheng2} study the caching gain by considering multiple antennas deployed at both transmit and receive nodes and show that the spatial multiplexing gain induced by multiple antennas and the caching gain are cumulative. Later on, the impact of caching has been studied in various wireless network models, such as combination networks \cite{combination_network1,combination_network2}, device-to-device networks \cite{d2d_twc1,d2d_twc2}, and fog radio access networks (Fog-RANs) \cite{xu_fog_ran,simeone_fog_ran,yuan_fog_ran}.

Note that many of the previous works on wireless caching assume a fully connected network topology, i.e., all the transmitters can communicate with all the receivers over independent and identically distributed (i.i.d.) fading channels. This assumption largely simplifies the theoretical analysis but also limits the applicability of these analytical results in practical wireless environment with \emph{partial connectivity} (due to path loss, shadowing, signal blocking, etc). In \cite{xu_partial_journal}, the authors attempt to address this issue by studying the storage-latency tradeoff in a partially connected interference network. They find that coded multicast gain and transmitter coordination gain, which are previously obtained in fully connected networks, can also be exploited in the partially connected network. The work \cite{gunduz_partial} focuses on a partially connected Fog-RAN, where each user is served by some locally connected edge nodes, which are not equipped with cache and connected to a cloud server via dedicated fronthaul links. The above works \cite{xu_partial_journal,gunduz_partial}, however, are limited to the one-dimensional linear network only. Later on, the authors in \cite{partially_cellular} consider a two-dimensional cache-aided cellular network modeled by a hexagonal topology where each user can only be connected to three nearby BSs. They obtain an order-optimal per-cell DoF result and show that the per-cell DoF scales linearly with the total amount of cache available in the cell. The delivery scheme in \cite{partially_cellular} is restricted to linear one-shot processing, and only exploits zero-forcing gain by activating the BSs caching the same requested subfiles at each time. The authors in \cite{Xinping_topo} formulate a joint file placement and delivery optimization problem to investigate the storage-latency tradeoff in a network with given and arbitrary partial connectivity. They adopt a non-splitting cache placement scheme, and thus cannot exploit the traffic load balancing gain by file splitting. Moreover, there is no information-theoretical analysis for the optimality of their results.

In this work, we consider a downlink cache-aided cellular network where the locations of cache-aided BSs are modeled as a grid topology with each grid regarded as a cell\footnote{For analytical tractability, it is a common practice in the literature to model the locations of BSs as a regular form in a two-dimensional plane, such as grid topology \cite{grid_topology1,grid_topology2} and hexagonal topology \cite{partially_cellular,cellular}.}. Considering path loss and shadowing effect, we assume that each user within a grid cell can only communicate with the nearby four BSs located at the corners of the cell. To ensure that the whole database is entirely cached at any four BSs around one user, in this paper, we focus on the cache size region $\mu\geq\frac{1}{4}$ with $\mu$ denoting the fraction of the database that each EN can store locally. The goal of this paper is to investigate how much capacity gain can be achieved by caching in the cellular network. Towards this end, we characterize the DoF, the asymptotic capacity measure in high signal-to-noise (SNR) regime, of the cache-aided cellular network. The main contributions and findings of this paper are summarized as follows:

 $\bullet$ \emph{Achievable DoF}: We adopt a file splitting and placement scheme tailored for the cellular network with partial connectivity. According to the overlapped degree of cached contents among BSs, caching brings different levels of BS cooperation for content delivery. We propose transmission schemes with no BS cooperation, partial BS cooperation, and full BS cooperation, respectively, for three typical cache sizes. In specific, for cache size $\mu=\frac{1}{4}$ where there is no BS cooperation, each BS only emits message to one nearby user at each time and interference alignment is used to coordinate signals of distinct cached contents among BSs. For cache size $\mu=\frac{1}{2}$ where there exists partial BS cooperation, the main idea is to utilize the common cached contents to cancel part of interference via interference neutralization and align the rest by interference alignment. For cache size $\mu=1$ where the BSs can fully cooperate, the whole cellular network can be regarded as a virtual multiple-input single-output (MISO) broadcast channel with partial connectivity and all the interference can be neutralized at each user by using a zero-forcing precoding. By applying memory sharing, we obtain the achievable DoF results for the whole cache size region $\mu\in[\frac{1}{4},1]$. Our achievable results reveal that the reciprocal of per-user DoF of the cellular network decreases piecewise linearly with the cache size $\mu$, and the gain of BS caching is more significant for the small cache size region.

 $\bullet$ \emph{Converse}: We first derive an outer bound on the DoF region of the cache-aided cellular network for any given uncoded prefetching scheme by using a genie-aided approach. Note that this outer bound is applicable not only for the cellular network with grid topology considered in this paper, but also for the cellular networks with other general topologies. Then, under the adopted cache placement scheme, we obtain an upper bound of per-user DoF for the considered cache-aided cellular network. It is shown that the obtained per-user DoF is optimal when $\mu\in\left[\frac{1}{2},1\right]$, and within an additive gap of $\frac{4}{39}$ to the optimum when $\mu\in\left[\frac{1}{4},\frac{1}{2}\right)$.

The remainder of the paper is organized as follows. Section II provides the system model of the cache-aided wireless cellular network and the adopted cache placement scheme. In Section III, we present the achievable DoF results and the corresponding proofs. The converse is given in Section IV. Section V concludes this paper. The proofs of some lemmas and theorems are relegated to the appendices.

Notations: $x$, ${\bf x}$, ${\bf X}$ and $\mathcal{X}$ denote scalar, vector, matrix and set, respectively.  ${\bf X}^{\dag}$ denotes the Moore-Penrose inverse of matrix ${\bf X}$. $\text{span}({\bf X})$ stands for the column span space of the matrix ${\bf X}$. $|\mathcal{X}|$ denotes the size of the set $\mathcal{X}$. $[n]$ denotes the set $\{1,2,\cdots,n\}$ where $n$ is an integer. $\{x^m\}_{m\in[n]}$ (or $\{x_m\}_{m\in[n]}$) denotes the set $\{x^1,x^2,\cdots, x^n\}$. ${\mathcal E}=\{0,\pm 2,\pm 4,\cdots\}$ denotes the set of all even numbers. ${\mathcal O}=\{\pm 1,\pm 3,\pm 5,\cdots\}$ denotes the set of all odd numbers. The important notations used in this paper are summarized in Table \ref{table}.

\begin{table}
 \centering
 \caption{List of Important Notations}\label{table}
   \begin{tabular}{|c|c|}
     \hline
     $(i,j)$ with $i,j\in{\mathcal O}$ & Coordinate of a user  \\
     \hline
     $(p,q)$ with $p,q\in{\mathcal E}$ & Coordinate of a BS \\
     \hline
     $\Psi$ & Set of all users  \\
     \hline
     $\Phi$ & Set of all BSs \\
     \hline
     ${\mathcal U}_1,{\mathcal U}_2,{\mathcal U}_3,{\mathcal U}_4$ & Four specific user subsets\\
     \hline
     ${\mathcal B}_1,{\mathcal B}_2,{\mathcal B}_3,{\mathcal B}_4$ & Four specific BS subsets \\
     \hline
     ${\mathcal M}_{(i,j)}$   & Four nearby BSs around user $(i,j)$  \\
     \hline
      $L$ & Number of files in the database \\
     \hline
    $\Omega_{(p,q)}$   & Cached contents of BS $(p,q)$  \\
     \hline
     $\mathcal A_k$ & BS cache group $k$\\
     \hline
     $\Theta $   & Set of all BS cache groups \\
     \hline
       $T$   & Codeword length \\
     \hline
     $ N $   & Symbol extension number \\
     \hline
     $ M $   & Signal Dimension \\
     \hline
   \end{tabular}
 \end{table}

\section{System Model}
\subsection{Cellular Network Model}\label{system_model}
\begin{figure}[t]
    \begin{centering}
  \includegraphics[scale=0.75]{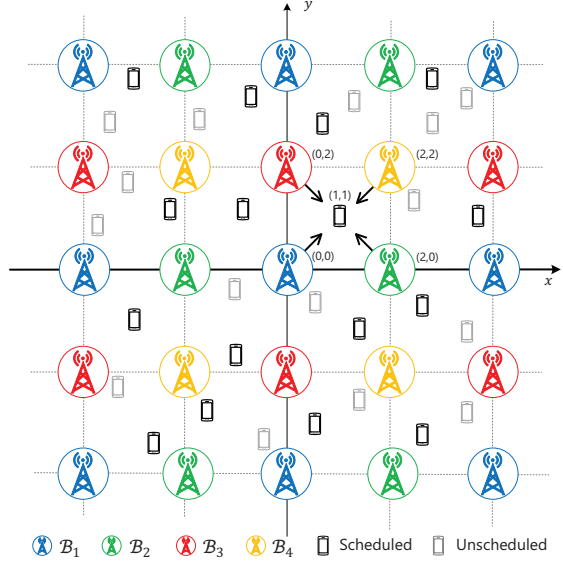}
   \caption{Cellular network model. Each user can only communicate with the four nearby BSs located at the corners of the grid cell. In each scheduling interval, one user with each grid cell is activated to be served.} \label{model}
   \end{centering}
    \end{figure}

We consider a cellular network where the BSs are located evenly on a two-dimensional plane, modeled as a square grid shown in Fig. \ref{model}. We use coordinate $(p,q)$, with $p,q\in{\mathcal E}$, to denote a BS. Each square formed by every set of four neighboring BSs is regarded as a cell.
The users are arbitrarily distributed. The user density is assumed to be large enough so that we can always schedule one (and at most one) user in each cell to serve in each resource block. The resource blocks can be defined in either time domain by using TDMA or frequency domain by using OFDMA. Without loss of generality, we consider the transmission in one resource block during one scheduling interval. We use coordinate $(i,j)$, with $i,j\in{\mathcal O}$, to denote the scheduled user within each cell. Considering the shadowing effect and distance-dependent path loss, we assume that the user within each cell can only be connected to the four nearby BSs located at the corners of the cell without interference from the other BSs. This assumption is valid since each mobile user only reports channel quality indicator (CQI) feedback from at most $3\sim4$ dominant nearby BSs in 3GPP standards \cite{partially_cellular}. Let ${\mathcal M}_{(i,j)}$ denote the set of the four nearby BSs around user $(i,j)$, i.e.
 \begin{align}
   {\mathcal M}_{(i,j)}\triangleq\left\{(i\pm1,j\pm1)\right\}.
 \end{align}
 Define $\Phi$ and $\Psi$ as the sets of all BSs and all users respectively.\footnote{We consider a large-scale system, where the number of BSs $|\Phi|$ and the number of users $|\Psi|$ are large enough, such that the effect of network margin can be ignored. } We further divide the set of all BSs $\Phi$ into four disjoint sets according to their geometrical coordinates as follows:
 \begin{subequations}\label{bs_set}
  \begin{align}
  {\mathcal B}_1&=\left\{(p,q):p=4z_1,q=4z_2,  z_1,z_2 \in \mathbb{Z}\right\}, \\
  {\mathcal B}_2&=\left\{(p,q):p=4z_1+2,q=4z_2,  z_1,z_2 \in \mathbb{Z}\right\}, \\
  {\mathcal B}_3&=\left\{(p,q):p=4z_1,q=4z_2+2,  z_1,z_2 \in \mathbb{Z}\right\}, \\
  {\mathcal B}_4&=\left\{(p,q):p=4z_1+2,q=4z_2+2,  z_1,z_2 \in \mathbb{Z}\right\},
\end{align}
\end{subequations}
which are demonstrated with different colors as shown in Fig. \ref{model}. These BS sets shall be used to design cache placement as discussed shortly later.

At each time slot $t$, the received signal at user $(i,j)$ can be represented as
\begin{align}
y_{(i,j)}(t)=\sum\limits_{(p,q)\in{\mathcal M}_{(i,j)}}h_{(i,j),(p,q)}(t)x_{(p,q)}(t)+n_{(i,j)}(t).
\end{align}
Here $h_{(i,j),(p,q)}(t)$ denotes the channel coefficient from BS $(p,q)$ to user $(i,j)$, which is time-variant and i.i.d. from a continuous distribution, $x_{(p,q)}(t)$ is the transmitted signal from BS $(p,q)$ subject to an average power constraint $P$, and $n_{(i,j)}(t)$ denotes the additive white Gaussian noise (AWGN) at user $(i,j)$ with zero mean and unit variance. Following
the convention in \cite{Ali2,xu,Ali3,xu_partial_journal,gunduz_partial,partially_cellular}, we assume that the perfect channel state information is available at all BSs and all users.

\subsection{Cache Model}\label{cache_model}
Consider a database consisting of $L$ files, denoted by $\{W_1,W_2,\cdots,W_L\}$, each of size $F$ bits. Each BS is equipped with a local cache and can cache up to $QF$ bits with $Q\leq L$. The normalized cache size is defined as $\mu\triangleq\frac{Q}{L}$, which represents the fraction of the database that each BS can store locally. Considering that in practical scenario, the memory equipped at a mobile user is negligible compared to the size of the whole database, we do not take the user-side cache into consideration. To ensure that the database can be entirely cached at any four BSs around one user, in this paper, we focus on cache size region $\mu\in\left[\frac{1}{4},1\right]$.

The system operates in two phases, a \emph{cache placement phase} and a \emph{content delivery phase}. In the cache placement phase, each BS $(p,q)$ maps the $L$ files in the database to its local cached contents $\Omega_{(p,q)}$ subject to its cache size constraint without knowing the future user demands. In this paper, we do not allow coding during the cache placement phase. A set of BSs ${\mathcal A}$ satisfying
\begin{subequations}
\begin{align}
  &\hspace{20mm}\bigcap_{(p,q)\in{\mathcal A}}\Omega_{(p,q)}\neq\O, \\
  &\left(\bigcap_{(p,q)\in{\mathcal A}}\Omega_{(p,q)}\right)\bigcap\left(\bigcup_{(p',q')\in\Phi\setminus{\mathcal A}}\Omega_{(p',q')}\right)=\O,
\end{align}
\end{subequations}
is defined as a \emph{BS cache group}, which indicates the set of BSs caching some common contents that are not available in any other BSs. Denote the set of all BS cache groups as $\Theta\triangleq\{\mathcal A\}$.
We apply the cache placement scheme proposed in \cite{Alilimits} by taking the partial connectivity of the network into consideration. Based on the BS sets $\{{\mathcal B}_k\}_{k\in[4]}$ defined in \eqref{bs_set}, we present the cache placement schemes at three cache sizes $\mu\in\{\frac{1}{4},\frac{1}{2},1\}$, respectively. The cache placement scheme for other cache sizes $\mu\in[\frac{1}{4},1]$ can be designed by using memory sharing \cite{Alilimits}.
\begin{itemize}
  \item $\mu=\frac{1}{4}$: Each file $W_{\ell}$, for $\ell\in[L]$, is evenly partitioned into four subfiles $\{W_{\ell}^k\}_{k\in[4]}$ with size $\frac{1}{4}F$ bits, and each $W_{\ell}^k$ is cached in all the BSs of set ${\mathcal B}_k$. By doing this, the BS cache group of subfiles $\{W_{\ell}^k\}_{\ell\in[L]}$ is given by ${\mathcal A}_k ={\mathcal B}_k$, for $k\in [4]$. Clearly, there is no overlapping in cached contents among BSs ${\mathcal M}_{(i,j)}$.
  \item $\mu=\frac{1}{2}$: Each file $W_{\ell}$, for $\ell\in[L]$, is evenly split into six subfiles $\{W_{\ell}^k\}_{k\in[6]}$ with size $\frac{1}{6}F$ bits, and each is cached in a unique collection of two out of the four BS sets $\{{\mathcal B}_k\}_{k\in[4]}$. More specifically, the BS cache group of subfiles $\{W_{\ell}^k\}_{\ell\in[L]}$, for $k\in[6]$, is given by ${\mathcal A}_1={\mathcal B}_1\cup{\mathcal B}_2$, ${\mathcal A}_2={\mathcal B}_3\cup{\mathcal B}_4$, ${\mathcal A}_3={\mathcal B}_1\cup{\mathcal B}_3$, ${\mathcal A}_4={\mathcal B}_2\cup{\mathcal B}_4$, ${\mathcal A}_5={\mathcal B}_1\cup{\mathcal B}_4$ and ${\mathcal A}_6={\mathcal B}_2\cup{\mathcal B}_3$, respectively. There exists partial overlapping in cached contents among BSs ${\mathcal M}_{(i,j)}$.
  \item $\mu=1$: In this case, the cache size of each BS is large enough to store the whole database. There is only one BS cache group, i.e., the set of all BSs $\Phi$.
\end{itemize}

In the content delivery phase, each user $(i,j)$ requests a file $W_{r_{(i,j)}}$ from the database, where $r_{(i,j)}\in[L]$. We define ${\bf r}\triangleq(r_{(i,j)})_{(i,j)\in\Psi}$ as the vector of all user demands. Upon receiving the user demands $\bf r$, BSs encode the requested files into signal sequences by using a codeword with length $T$, and transmit them to satisfy user demands. The error probability of the system is defined as $P_e=\max\limits_{\bf r}\max\limits_{(i,j)\in\Psi} \mathbb{P}({\hat W}_{(i,j)}\neq W_{r_{(i,j)}})$, where ${\hat W}_{(i,j)}$ is the estimated file of user $(i,j)$ from the received signal sequence. Since each user receives a file with size $F$ bits during codeword length $T$, we define $R(\mu, P)\triangleq\frac{F}{T}$ as the per-user rate at cache size $\mu$ and transmit power $P$. The per-user rate $R(\mu, P)$ of the system is said to be achievable if and only if there exists a coding scheme that all users can decode their requested files with vanishing error probability, i.e., $P_e \rightarrow 0$ as $F\rightarrow \infty$.

\subsection{Performance Metric}
We define capacity $C(\mu,P)$ as the supremum of all achievable per-user rates for the given cache size and the power constraint. The per-user DoF of the system is defined as
\begin{align}
  d(\mu)\triangleq\lim\limits_{P\rightarrow \infty}\frac{C(\mu,P)}{\log P}.
\end{align}
Since the reciprocal of $d(\mu)$ is a convex function of cache size $\mu$ \cite{Ali2}, we will use $1/d$ to show our results.

\section{Achievable DoF}\label{achievable_part}
This section presents the achievable per-user DoF results for the cache-aided cellular network.

\begin{theorem}\label{theorem1}
For the considered cache-aided cellular network where each BS has a cache of normalized size $\mu$, the reciprocal of the per-user DoF $d$ is upper bounded as
\begin{align}
  1/d\leq1/d_{\text{lower}}\triangleq\begin{cases}
  \frac{17}{6}-\frac{10}{3}\mu, & \mu\in\left[\frac{1}{4},\frac{1}{2}\right), \\
  \frac{4}{3}-\frac{1}{3}\mu , & \mu\in\left[\frac{1}{2},1\right]. \\
  \end{cases}
\end{align}
\end{theorem}
\begin{remark}
  It can be seen that the reciprocal of the obtained per-user DoF $d$ decreases piecewise linearly with the normalized cache size $\mu$. In addition, the line segment at the small cache size region $\mu<\frac{1}{2}$ is steeper than that at $\mu\geq\frac{1}{2}$, indicating that the gain of BS caching is more significant at the small cache region.
\end{remark}

In the delivery phase, we consider the worst-case scenario where all user demands are distinct. When user demands are not all distinct, the same delivery strategy can be applied by treating the demands as being different. In the following three subsections, we present the achievable schemes of three corner points $(\mu=1/4,1/d=2)$, $(\mu=1/2,1/d=7/6)$ and $(\mu=1,1/d=1)$\footnote{The proposed achievable schemes of $\mu=\frac{1}{4}$ and $1$ are still applicable if the connectivity degree of each user is larger than 4. For instance, consider that each user can be connected to nearby 12 BSs, i.e., ${\mathcal M}'_{(i,j)}\triangleq\left\{(i\pm1,j\pm1), (i\pm3,j\pm1),(i\pm1,j\pm3)\right\}$. In this case, the delivery scheme of $\mu=\frac{1}{4}$ is very similar to Section \ref{no_cooperation} by taking ${\mathcal M}'_{(i,j)}$ into the construction of precoding matrices, and the achievable per-user DoF is $d=\frac{1}{2}$. The delivery of $\mu=1$ in this scenario also forms a virtual MISO broadcast channel, and per-user DoF $d=1$ is achieved by a zero-forcing precoding.}. The achievability of other points can be proved by using memory sharing \cite{Alilimits}.

\subsection{$\mu=\frac{1}{4}$ (No BS Cooperation)}\label{no_cooperation}
 By the cache placement scheme in Section \ref{cache_model}, the four requested subfiles of each user $(i,j)$ are cached in the four BS cache groups $\{{\mathcal A}_k\}_{k\in[4]}$, denoted as $\{W_{r_{(i,j)},{\mathcal A}_k}\}_{k\in[4]}$ respectively. Since there is no overlapping in the cached contents, no transmission cooperation can be exploited among the BS set ${\mathcal M}_{(i,j)}$. The main idea of the achievable scheme is to divide the whole transmission into different phases so that each BS only serves one user in each phase, and asymptotic interference alignment is used to align the interference for each user.

  \begin{figure}[t]
    \begin{centering}
  \includegraphics[scale=1.0]{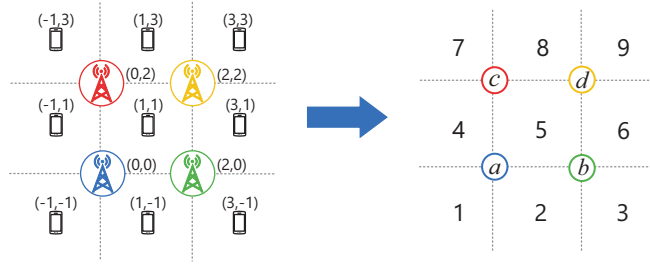}
   \caption{An example network consisting of some specific BSs and users. To simplify notation, we denote the users $(-1,-1)$, $(1,-1)$, $(3,-1)$, $(-1,1)$, $(1,1)$, $(3,1)$, $(-1,3)$, $(1,3)$, and $(3,3)$ as $1,2,3,4,5,6,7,8$, and $9$ respectively, and denote the BSs $(0,0)$, $(2,0)$, $(0,2)$, and $(2,2)$ as $a$, $b$, $c$, and $d$ respectively.} \label{simplified}
   \end{centering}
   \end{figure}

 We choose a part of the cellular network consisting of some specific BSs and users as an example shown in Fig. \ref{simplified} to present the achievable scheme, and the proof for the entire cellular network is placed in Appendix \ref{appendix_no_cooperation}. Here, the four BS cache groups are ${\mathcal A}_1=\{a\}$, ${\mathcal A}_2=\{b\}$, ${\mathcal A}_3=\{c\}$, and ${\mathcal A}_4=\{d\}$. Note that only the four BSs around user 5 are entirely given in this example network. In the rest of this subsection, we only present the specific transmission design associated with user 5 and show that it can achieve the target DoF. The achievable scheme is based on the asymptotic interference alignment \cite{cadambe2009xchannel}. In specific, we use \cite[Lemma 2]{cadambe2009xchannel} to design precoding matrices and the proof for the decodability of desired signals relies on \cite[Lemma 1]{cadambe2009xchannel}. However, due to the partial connectivity of the network, the space design of interference and the specific precoding construction in this paper are different from those in \cite{cadambe2009xchannel} which focuses on a fully connected interference network.

\begin{figure*}[t]
\begin{centering}
\includegraphics[scale=0.7]{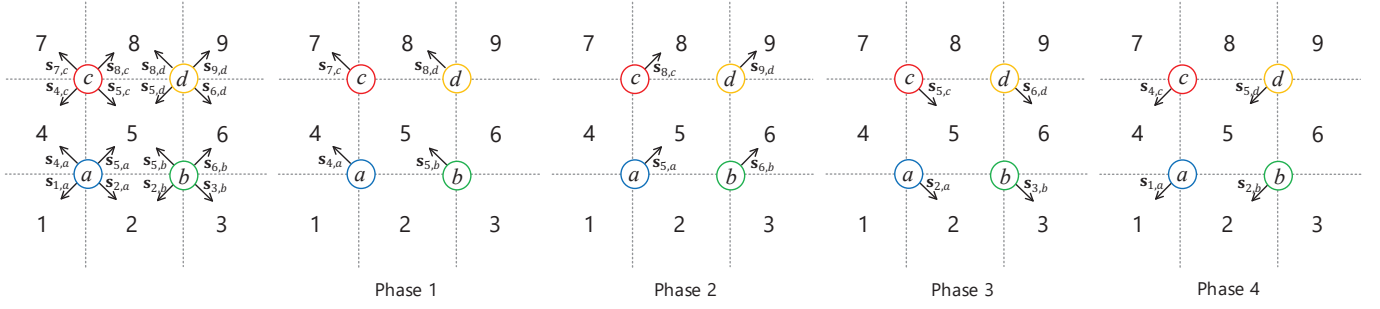}
 \caption{The four-phase alternating transmission scheme for $\mu=\frac{1}{4}$. In each phase, each BS only sends a signal to serve one of its nearby users. }\label{alternating}
\end{centering}
\end{figure*}

Each BS employs a random Gaussian coding scheme to encode the four subfiles requested by its serving users into signals, i.e.
\begin{equation*}
\begin{array}{lll}
  \text{BS } a : & W_{r_1,{\mathcal A}_1}\rightarrow {\bf s}_{1,a}, & W_{r_2,{\mathcal A}_1}\rightarrow {\bf s}_{2,a}, \\
  \              & W_{r_4,{\mathcal A}_1}\rightarrow {\bf s}_{4,a}, & W_{r_5,{\mathcal A}_1}\rightarrow {\bf s}_{5,a}, \\
  \text{BS } b : & W_{r_2,{\mathcal A}_2}\rightarrow {\bf s}_{2,b}, & W_{r_3,{\mathcal A}_2}\rightarrow {\bf s}_{3,b}, \\
  \              & W_{r_5,{\mathcal A}_2}\rightarrow {\bf s}_{5,b}, & W_{r_6,{\mathcal A}_2}\rightarrow {\bf s}_{6,b}, \\
  \text{BS } c : & W_{r_4,{\mathcal A}_3}\rightarrow {\bf s}_{4,c}, & W_{r_5,{\mathcal A}_3}\rightarrow {\bf s}_{5,c}, \\
  \              & W_{r_7,{\mathcal A}_3}\rightarrow {\bf s}_{7,c}, & W_{r_8,{\mathcal A}_3}\rightarrow {\bf s}_{8,c}, \\
  \text{BS } d : & W_{r_5,{\mathcal A}_4}\rightarrow {\bf s}_{5,d}, & W_{r_6,{\mathcal A}_4}\rightarrow {\bf s}_{6,d}, \\
  \              & W_{r_8,{\mathcal A}_4}\rightarrow {\bf s}_{8,d}, & W_{r_9,{\mathcal A}_4}\rightarrow {\bf s}_{9,d},
\end{array}
\end{equation*}
where each ${\bf s}_{(i,j),(p,q)}=[s^1_{(i,j),(p,q)},\cdots,s^M_{(i,j),(p,q)}]^T\in{\mathbb C}^M$ is the signal vector transmitted from BS $(p,q)$ and intended to user $(i,j)$. The whole transmission contains four phases and each phase takes $N$ symbol extensions. In each phase, each BS transmits one of the above four signals to serve one user as illustrated in Fig. \ref{alternating}. Without loss of generality, we take phase 1 as an example and show that the DoF of $\frac{1}{2}$ is achievable for user 5. Denote ${\bf v}^m_{(i,j),(p,q)}\in{\mathbb C}^{N\times 1}$ as the precoding vector of signal $s^m_{(i,j),(p,q)}$ with $m\in[M]$ and denote ${\bf V}_{(i,j),(p,q)}=[\begin{array}{ccc}
                                               {\bf v}^1_{(i,j),(p,q)} & \cdots & {\bf v}^M_{(i,j),(p,q)}
                                             \end{array}
  ]$. The choice of the signal dimension $M$ and symbol extension number $N$ is given shortly later in Remark \ref{dimension design for mu=1/4}. The received signal (ignoring noise) of user 5, denoted by ${\bf y}_5\in{\mathbb C}^{N}$, can be represented as
\begin{align}
  {\bf y}_{5}=\underbrace{{\bf H}_{5,b}{\bf V}_{5,b}{\bf s}_{5,b}}_{\text{desired signal}}+\underbrace{{\bf H}_{5,a}{\bf V}_{4,a}{\bf s}_{4,a}+{\bf H}_{5,c}{\bf V}_{7,c}{\bf s}_{7,c}+{\bf H}_{5,d}{\bf V}_{8,d}{\bf s}_{8,d}}_{\text{interference}},
\end{align}
where each ${\bf H}_{(i,j),(p,q)}\in{\mathbb C}^{N\times N}$ is the diagonal channel matrix from BS $(p,q)$ to user $(i,j)$ with elements $\{h_{(i,j),(p,q)}(\eta)\}_{\eta\in[N]}$. We aim to align the directions of the three interference signals into a common space, i.e.
\begin{subequations}\label{exam_align}
\begin{align}
  \textrm{span}\left({\bf H}_{5,a}{\bf V}_{4,a}\right)\subset \textrm{span}\left({\bf V}'\right), \\
  \textrm{span}\left({\bf H}_{5,c}{\bf V}_{7,c}\right)\subset \textrm{span}\left({\bf V}'\right), \\
  \textrm{span}\left({\bf H}_{5,d}{\bf V}_{8,d}\right)\subset \textrm{span}\left({\bf V}'\right),
\end{align}
\end{subequations}
where ${\bf V}'$ is an $N\times(N-M)$ matrix. Based on \cite[Lemma 2]{cadambe2009xchannel}, we shall design every element in the $\eta$-th row of ${\bf V}_{(i,j),(p,q)}$ as a multivariate monomial function of entries in the $\eta$-th rows of all interference channel matrices, which are denoted by a set ${\mathcal I}$. In this example, ${\mathcal I}$ is given by
\begin{align}
  {\mathcal I}=\{{\bf H}_{5,a},{\bf H}_{5,c},{\bf H}_{5,d}\}.
\end{align}
We construct the following set of column vectors
\begin{align}\label{set_v(n)}
{\mathcal V}(n)=\big\{(\left({\bf H}_{5,a}\right)^{\iota_{5,a}}\left({\bf H}_{5,c}\right)^{\iota_{5,c}}({\bf H}_{5,d})^{\iota_{5,d}}{\bf 1}_{N}): \iota_{5,a},\iota_{5,c},\iota_{5,d}\in[n]\big\}.
\end{align}
 where ${\bf 1}_{N}$ is an $N\times 1 $ vector with each element being one. The column vectors in ${\mathcal V}(n)$ can occur in any order to form each of matrices ${\bf V}_{5,b}$ ${\bf V}_{4,a}$, ${\bf V}_{7,c}$ and ${\bf V}_{8,d}$. For example, the $m$-th column of ${\bf V}_{5,b}$, ${\bf v}^m_{5,b}$ can be expressed as
\begin{align}
  {\bf v}^m_{5,b}=\left({\bf H}_{5,a}\right)^{\iota_{5,a}^m}\left({\bf H}_{5,c}\right)^{\iota_{5,c}^m}\left({\bf H}_{5,d}\right)^{\iota_{5,d}^m}{\bf 1}_{N},
\end{align}
where the subscripts of each exponent $\iota_{(i,j),(p,q)}^{m}\in[n]$ are associated
with corresponding base ${\bf H}_{(i,j),(p,q)}$, and its superscript $m$ denotes the column index of ${\bf v}^m_{5,b}$ in ${\bf V}_{5,b}$. The received signal at user $5$ can be rewritten as
\begin{align}
  {\bf y}_{5}=&\underbrace{{\bf H}_{5,b}\sum\limits_{m=1}^{M}\left({\bf H}_{5,a}\right)^{\iota_{5,a}^m}\left({\bf H}_{5,c}\right)^{\iota_{5,c}^m}\left({\bf H}_{5,d}\right)^{\iota_{5,d}^m}{\bf 1}_{N}s_{5,b}^{m}}_{\text{desired signal}}\nonumber\\
  &+\underbrace{{\bf H}_{5,a}\sum\limits_{m=1}^{M}\left({\bf H}_{5,a}\right)^{\iota_{5,a}^m}\left({\bf H}_{5,c}\right)^{\iota_{5,c}^m}\left({\bf H}_{5,d}\right)^{\iota_{5,d}^m}{\bf 1}_{N}s_{4,a}^{m}}_{\text{interference}}\nonumber \\
  &+\underbrace{{\bf H}_{5,c}\sum\limits_{m=1}^{M}\left({\bf H}_{5,a}\right)^{\iota_{5,a}^m}\left({\bf H}_{5,c}\right)^{\iota_{5,c}^m}\left({\bf H}_{5,d}\right)^{\iota_{5,d}^m}{\bf 1}_{N}s_{7,c}^{m}}_{\text{interference}}\nonumber\\
  &+\underbrace{{\bf H}_{5,d}\sum\limits_{m=1}^{M}\left({\bf H}_{5,a}\right)^{\iota_{5,a}^m}\left({\bf H}_{5,c}\right)^{\iota_{5,c}^m}\left({\bf H}_{5,d}\right)^{\iota_{5,d}^m}{\bf 1}_{N}s_{8,d}^{m}}_{\text{interference}}.
\end{align}
 Note that each precoding vector of interference after multiplying channel matrix has a similar monomial structure as that in ${\mathcal V}(n)$, except that the range of each exponent is changed from $[n]$ to $[n+1]$. This indicates that these received precoding vectors of interference belong to the set ${\mathcal V}(n+1)$. Thus, by letting the matrix ${\bf V}'$ consist of all the vectors in set ${\mathcal V}(n+1)$, the interference alignment conditions \eqref{exam_align} are satisfied.
\begin{remark}\label{dimension design for mu=1/4}
  Since each user is interfered by three signals in the delivery, the total number of interference channels ${\mathcal I}$ is $|\mathcal I|=3|\Psi|$ (in this example, $|\mathcal I|=3$ due to that we only consider three interference channels for user 5). Note that the columns of each precoding matrix ${\bf V}_{(i,j),(p,q)}$ are formed by all the vectors in set ${\mathcal V}(n)$ with any order. Thus, the signal dimension $M$ is equal to the cardinality of set ${\mathcal V}(n)$, i.e., $M=|{\mathcal V}(n)|=n^{3|\Psi|}$. On the other hand, the dimension of all received interference vectors is equal to the cardinality of set ${\mathcal V}(n+1)$ and given by $(n+1)^{3|\Psi|}$. To ensure the decodability, the symbol extension number should be $N=n^{3|\Psi|}+(n+1)^{3|\Psi|}$.
\end{remark}

Next, we show that for each user, the received vectors of all desired signals are linearly independent from those of interference. For user 5, we shall prove that the matrix
\begin{align}
  {\bf \Lambda}_{5}=\left[{\bf H}_{5,b}{\bf V}_{5,b}\quad {\bf V}'\right]
\end{align}
has full rank almost surely. To apply \cite[Lemma 1]{cadambe2009xchannel}, we need to verify that the elements in the same row of ${\bf \Lambda}_{5}$ are different monomials. We have the following observations:
\begin{enumerate}
  \item In the $\eta$-th row of ${\bf \Lambda}_{5}$, the products in ${\bf H}_{5,b}{\bf V}_{5,b}$ are different from each other because each column ${\bf v}^m_{5,b}$ with $m\in[M]$ is a unique vector in set ${\mathcal V}(n)$; Similarly, the terms in the $\eta$-th row of ${\bf V}'$ are also different from each other since each column of ${\bf V}'$ is unique in set ${\mathcal V}(n+1)$.
  \item Since ${\bf H}_{5,b}$ is not in the construction of ${\mathcal V}(n)$, the terms in the $\eta$-th row of ${\bf H}_{5,b}{\bf V}_{5,b}$ and ${\bf V}'$ differ in channel factor $h_{5,b}(\eta)$.
\end{enumerate}
Therefore, the condition of \cite[Lemma 1]{cadambe2009xchannel} is met and the matrix ${\bf \Lambda}_{5}$ has a full rank almost surely. Since user 5 can decode $n^{|3\Psi|}$ desired signals over $n^{3|\Psi|}+(n+1)^{3|\Psi|}$ symbol extensions, the achievable per-user DoF can be expressed as
\begin{align}
  d=\frac{n^{|3\Psi|}}{n^{3|\Psi|}+(n+1)^{3|\Psi|}}.
\end{align}
Note that the above per-user $d$ increases monotonically with $n$. Taking $n\rightarrow\infty$, we can achieve the maximum value $\frac{1}{2}$ of the per-user DoF asymptotically.

\subsection{$\mu=\frac{1}{2}$ (Partial BS Cooperation)}\label{partial_cooperation}
By the cache placement in Section \ref{cache_model}, the six requested subfiles of user $(i,j)$ are cached in $\{{\mathcal A}_k\}_{k\in[6]}$, respectively. There exists partial overlapping in the cached contents among BSs ${\mathcal M}_{(i,j)}$. Thus, partial transmission cooperation can be exploited among BSs ${\mathcal M}_{(i,j)}$. In this subsection, the main idea of the achievable scheme for the cellular network with partial BS cooperation is to first cancel part of the undesired signals and then align the rest of interference for each user by using asymptotic interference alignment. In this case, BSs in each BS cache group ${\mathcal A}_k$ use a random Gaussian coding scheme to encode subfile $W_{r_{(i,j)},{\mathcal A}_k}$ requested by user $(i,j)$ into signal ${\bf s}_{(i,j),{\mathcal A}_k}\in{\mathbb C}^M$. Denote ${\bf V}_{(i,j),{\mathcal A}_k}^{(p,q)}\in\mathbb{C}^{N\times M}$ as the precoding matrix of signal ${\bf s}_{(i,j),{\mathcal A}_k}$ at BS $(p,q)$ where $(p,q)\in{{\mathcal A}_k}$. (The signal dimension $M$ and symbol extension number $N$ will be determined shortly in Remark \ref{dimension design for mu=1/2}.) Each precoding matrix ${\bf V}_{(i,j),{\mathcal A}_k}^{(p,q)}$ is a product of two matrices, i.e.
\begin{equation}
{\bf V}_{(i,j),{\mathcal A}_k}^{(p,q)}={\bf U}_{(i,j),{\mathcal A}_k}^{(p,q)}{\bf W}_{(i,j),{\mathcal A}_k},
\end{equation}
where ${\bf U}_{(i,j),{\mathcal A}_k}^{(p,q)}$ is an $N\times N$ diagonal matrix and ${\bf W}_{(i,j),{\mathcal A}_k}=[\begin{array}{cccc}
{\bf w}^{1}_{(i,j),{\mathcal A}_k} & {\bf w}^{2}_{(i,j),{\mathcal A}_k} & \cdots & {\bf w}^{M}_{(i,j),{\mathcal A}_k}\end{array}]$
 is an $N\times M$ matrix. The above product form of ${\bf V}_{(i,j),{\mathcal A}_k}^{(p,q)}$ is inspired by the previous work\cite{xu} on cooperative X-multicast channels. The first matrix ${\bf U}_{(i,j),{\mathcal A}_k}^{(p,q)}$ is designed to cancel part of the interference while the second matrix ${\bf W}_{(i,j),{\mathcal A}_k}$ is used to apply asymptotic interference alignment. Similar to Section \ref{no_cooperation}, we use the example network shown in Fig. \ref{simplified} to illustrate the design of matrices ${\bf U}_{(i,j),{\mathcal A}_k}^{(p,q)}$ and ${\bf W}_{(i,j),{\mathcal A}_k}$. The specific design methods of ${\bf U}_{(i,j),{\mathcal A}_k}^{(p,q)}$ and ${\bf W}_{(i,j),{\mathcal A}_k}$ for the entire cellular network are given in Appendix \ref{appendix_partial_cooperation}. In this case, the BS cache groups in Fig. \ref{simplified} are ${\mathcal A}_1=\{a,b\}$, ${\mathcal A}_2=\{c,d\}$, ${\mathcal A}_3=\{a,c\}$, ${\mathcal A}_4=\{b,d\}$, ${\mathcal A}_5=\{a,d\}$ and ${\mathcal A}_6=\{b,c\}$.

 We first elaborate the design of ${\bf U}_{(i,j),{\mathcal A}_k}^{(p,q)}$. For notation convenience, we define
 \begin{align}
{\bf G}^{(i',j')}_{(i,j),{\mathcal A}_k}\triangleq{\bf H}_{(i,j),(p,q)}{\bf U}_{(i',j'),{\mathcal A}_k}^{(p,q)}+{\bf H}_{(i,j),(p',q')}{\bf U}_{(i',j'),{\mathcal A}_k}^{(p',q')},\quad \text{with } \{(p,q),(p',q')\}={\mathcal M}_{(i,j)}\cap{\mathcal A}_k,
 \end{align}
 as the effective channel of interference at user $(i,j)$, which is caused by the signal transmitted from BS cache group ${\mathcal A}_k$ to user $(i',j')$. The design of ${\bf U}_{(i,j),{\mathcal A}_k}^{(p,q)}$ contains the following three forms:

\begin{itemize}
  \item Case I: ${\bf U}_{(i,j),{\mathcal A}_k}^{(p,q)}$ is designed as a function of channel matrices. Since some adjacent BSs are connected to two common users, e.g., both BSs $a$ and $b$ can simultaneously serve users $2$ and $5$, we can use zero-forcing precoding to neutralize interference.  We take signal ${\bf s}_{2,{\mathcal A}_1}$ for an example, which is transmitted cooperatively from BSs $a,b\in{\mathcal A}_1$ and interferes user $5$. By designing ${\bf U}_{2,{\mathcal A}_1}^{a}={\bf H}_{5,b}$ and ${\bf U}_{2,{\mathcal A}_1}^{b}=-{\bf H}_{5,a}$, for user $5$, the interference caused by signal ${\bf s}_{2,{\mathcal A}_1}$ is neutralized as
     \begin{align}
      {\bf G}^{2}_{5,{\mathcal A}_1}{\bf W}_{2,{\mathcal A}_1}{\bf s}_{2,{\mathcal A}_1}=({\bf H}_{5,a}{\bf H}_{5,b}-{\bf H}_{5,b}{\bf H}_{5,a}){\bf W}_{2,{\mathcal A}_1}{\bf s}_{2,{\mathcal A}_1}={\bf 0}.
     \end{align}
     Similarly, for other adjacent BSs connected to two common users, the corresponding precoding matrices can also be designed by zero-forcing. For user 5, the effective channels of interference that can be neutralized as zeros are given by ${\bf G}^{2}_{5,{\mathcal A}_1}$, ${\bf G}^{8}_{5,{\mathcal A}_2}$, ${\bf G}^{4}_{5,{\mathcal A}_3}$ and ${\bf G}^{6}_{5,{\mathcal A}_4}$.

\item Case II: ${\bf U}_{(i,j),{\mathcal A}_k}^{(p,q)}$ is designed as a zero matrix. Due to the partial connectivity of the network, we can force some ${\bf U}_{(i,j),{\mathcal A}_k}^{(p,q)}$ with $(p,q)\notin{\mathcal M}_{(i,j)}$ to be zeros in order to avoid interference. For example, noticing that both BSs $a$ and $b$ cannot be connected to user $8$, i.e., $a,b\notin{\mathcal M}_8$, we can design  ${\bf U}^{a}_{8,{\mathcal A}_1}={\bf U}^{b}_{8,{\mathcal A}_1}={\bf 0}$ and the interference caused by signal ${\bf s}_{8,{\mathcal A}_1}$ at user 5 can be cancelled as
   \begin{align}
    &{\bf G}^{8}_{5,{\mathcal A}_1}{\bf W}_{8,{\mathcal A}_1}{\bf s}_{8,{\mathcal A}_1}=({\bf H}_{5,a}\cdot{\bf 0}+{\bf H}_{5,b}\cdot{\bf 0}){\bf W}_{8,{\mathcal A}_1}{\bf s}_{8,{\mathcal A}_1}={\bf 0}.
   \end{align}
    Similarly, for user 5, we can obtain the following effective channels of interference as zeros: ${\bf G}^{7}_{5,{\mathcal A}_1}$, ${\bf G}^{8}_{5,{\mathcal A}_1}$, ${\bf G}^{9}_{5,{\mathcal A}_1}$, ${\bf G}^{1}_{5,{\mathcal A}_2}$, ${\bf G}^{2}_{5,{\mathcal A}_2}$, ${\bf G}^{3}_{5,{\mathcal A}_2}$, ${\bf G}^{3}_{5,{\mathcal A}_3}$, ${\bf G}^{6}_{5,{\mathcal A}_3}$, ${\bf G}^{9}_{5,{\mathcal A}_3}$, ${\bf G}^{1}_{5,{\mathcal A}_4}$, ${\bf G}^{4}_{5,{\mathcal A}_4}$ and ${\bf G}^{7}_{5,{\mathcal A}_4}$. Note that not all $\{{\bf U}_{(i,j),{\mathcal A}_k}^{(p,q)}\}$ with $(p,q)\notin{\mathcal M}_{(i,j)}$ are forced to zeros. The remaining $\{{\bf U}_{(i,j),{\mathcal A}_k}^{(p,q)}\}$ are designed in case III.

\item Case III: Each diagonal element of ${\bf U}_{(i,j),{\mathcal A}_k}^{(p,q)}$ is i.i.d. from a continuous distribution. This design can make the polynomial functions of the entries in ${\bf G}^{(i',j')}_{(i,j),{\mathcal A}_k}$ linearly independent with those of the effective channels experienced by the desired signals. This condition will be used to prove the decodability of the desired signals and please see Appendix \ref{appendix_partial_cooperation} for more details.
\end{itemize}
 In summary, we give the nonzero effective channels of interference at user 5 from each BS cache group ${\mathcal A}_k$ with $k\in[6]$ as follows
\begin{subequations}
  \begin{align}
{\mathcal J}_{5,{\mathcal A}_1}=\Big\{&{\bf G}^{1}_{5,{\mathcal A}_1},{\bf G}^{3}_{5,{\mathcal A}_1},{\bf G}^{4}_{5,{\mathcal A}_1},{\bf G}^{6}_{5,{\mathcal A}_1}\Big\},&\\
{\mathcal J}_{5,{\mathcal A}_2}=\Big\{&{\bf G}^{4}_{5,{\mathcal A}_2},{\bf G}^{6}_{5,{\mathcal A}_2},{\bf G}^{7}_{5,{\mathcal A}_2},{\bf G}^{9}_{5,{\mathcal A}_2}\Big\},\\
{\mathcal J}_{5,{\mathcal A}_3}=\Big\{&{\bf G}^{1}_{5,{\mathcal A}_3},{\bf G}^{2}_{5,{\mathcal A}_3},{\bf G}^{7}_{5,{\mathcal A}_3},{\bf G}^{8}_{5,{\mathcal A}_3}\Big\},\\
{\mathcal J}_{5,{\mathcal A}_4}=\Big\{&{\bf G}^{2}_{5,{\mathcal A}_4},{\bf G}^{3}_{5,{\mathcal A}_4},{\bf G}^{8}_{5,{\mathcal A}_4},{\bf G}^{9}_{5,{\mathcal A}_4}\Big\},\\
{\mathcal J}_{5,{\mathcal A}_5}=\Big\{&{\bf G}^{1}_{5,{\mathcal A}_5},{\bf G}^{2}_{5,{\mathcal A}_5},{\bf G}^{3}_{5,{\mathcal A}_5},{\bf G}^{4}_{5,{\mathcal A}_5},{\bf G}^{6}_{5,{\mathcal A}_5},{\bf G}^{7}_{5,{\mathcal A}_5},{\bf G}^{8}_{5,{\mathcal A}_5},{\bf G}^{9}_{5,{\mathcal A}_5}\Big\},\\
{\mathcal J}_{5,{\mathcal A}_6}=\Big\{&{\bf G}^{1}_{5,{\mathcal A}_6},{\bf G}^{2}_{5,{\mathcal A}_6},{\bf G}^{3}_{5,{\mathcal A}_6},{\bf G}^{4}_{5,{\mathcal A}_6},{\bf G}^{6}_{5,{\mathcal A}_6},{\bf G}^{7}_{5,{\mathcal A}_6},{\bf G}^{8}_{5,{\mathcal A}_6},{\bf G}^{9}_{5,{\mathcal A}_6}\Big\}.
\end{align}
\end{subequations}
 The cardinality of each set ${\mathcal J}_{5,{\mathcal A}_k}$ with $k\in[4]$ is $2(\sqrt{|\Psi|}-1)=2(\sqrt{9}-1)=4$ and the cardinality of each set ${\mathcal J}_{5,{\mathcal A}_k}$ with $k\in\{5,6\}$ is $|\Psi|-1=9-1=8$. We further denote ${\mathcal J}_{(i,j)}$ as all the nonzero effective channels of interference at user $(i,j)$, e.g.
\begin{align}
  {\mathcal J}_{5}={\mathcal J}_{5,{\mathcal A}_1}\cup{\mathcal J}_{5,{\mathcal A}_2}\cup{\mathcal J}_{5,{\mathcal A}_3}\cup{\mathcal J}_{5,{\mathcal A}_4}\cup{\mathcal J}_{5,{\mathcal A}_5}\cup{\mathcal J}_{5,{\mathcal A}_6}.
\end{align}

\begin{figure*}[t]
    \begin{centering}
  \includegraphics[scale=0.7]{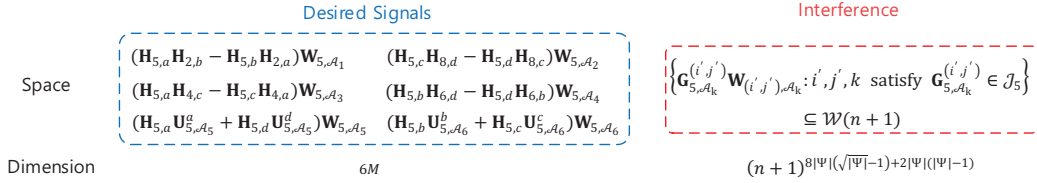}
   \caption{Signal space of the received signal at user $5$ in the example network.} \label{signal space}
   \end{centering}
   \end{figure*}

Then we present the design of ${\bf W}_{(i,j),{\mathcal A}_k}$. We aim to align the directions of interference at each user into a common space. For user 5, the interference alignment conditions can be expressed as
\begin{align} \label{interference alignment condtion 2}
 &\textrm{span}\left({\bf G}^{(i',j')}_{5,{\mathcal A}_k}{\bf W}_{(i',j'),{\mathcal A}_k}\right)\subset\textrm{span}\left({\bf W}'\right),\quad  \forall i',j',k\  \text{satisfy} \ {\bf G}^{(i',j')}_{5,{\mathcal A}_k}\in{\mathcal J}_5,
\end{align}
where ${\bf W}'$ is an $N\times(N-6M)$ matrix. Similar to Section \ref{no_cooperation}, we shall design every entry in the $\eta$-th row of ${\bf W}_{(i,j),{\mathcal A}_k}$ as a multivariate monomial function of entries in the $\eta$-th rows of effective channels of interference in ${\mathcal J}_5$. Towards this end, we construct the following set of vectors
\begin{align}\label{example_set_w(n)}
  {\mathcal W}(n)=\left\{\left(\prod\limits_{i',j',k: {\bf G}^{(i',j')}_{5,{\mathcal A}_k}\in {\mathcal J}_5
   }\Big({\bf G}^{(i',j')}_{5,{\mathcal A}_k}\Big)^{\iota_{(i',j'),k}}{\bf 1}_{N}\right):\iota_{(i',j'),k}\in[n]\right\}
\end{align}
where the subscripts of each exponent $\iota_{(i',j'),k}$ are associated with corresponding base ${\bf G}^{(i',j')}_{5,{\mathcal A}_k}$. The column vectors in ${\mathcal W}(n)$ can occur in any order to form each of $\{{\bf W}_{(i',j'),{\mathcal A}_k}: i',j',k\  \text{satisfy}$ ${\bf G}^{(i',j')}_{5,{\mathcal A}_k}\in{\mathcal J}_5\}$. In this way, the precoding vectors of the interference after multiplying channel matrices belong to the set ${\mathcal W}(n+1)$. The interference alignment conditions \eqref{interference alignment condtion 2} are satisfied by letting the matrix ${\bf W}'$ consist of all the vectors in set ${\mathcal W}(n+1)$. The received signal at user $5$ can be expressed as
\begin{align}
{\bf y}_{5}=&\left({\bf H}_{5,a}{\bf H}_{2,b}-{\bf H}_{5,b}{\bf H}_{2,a}\right){\bf W}_{5,{\mathcal A}_1}{\bf s}_{5,{\mathcal A}_1}\nonumber \\
            &+\left({\bf H}_{5,c}{\bf H}_{8,d}-{\bf H}_{5,d}{\bf H}_{8,c}\right){\bf W}_{5,{\mathcal A}_2}{\bf s}_{5,{\mathcal A}_2}\nonumber \\
            &+\left({\bf H}_{5,a}{\bf H}_{4,c}-{\bf H}_{5,c}{\bf H}_{4,a}\right){\bf W}_{5,{\mathcal A}_3}{\bf s}_{5,{\mathcal A}_3}\nonumber \\
            &+\left({\bf H}_{5,b}{\bf H}_{6,d}-{\bf H}_{5,d}{\bf H}_{6,b}\right){\bf W}_{5,{\mathcal A}_4}{\bf s}_{5,{\mathcal A}_4} \nonumber \\
            &+\left({\bf H}_{5,a}{\bf U}_{5,{\mathcal A}_5}^{a}+{\bf H}_{5,d}{\bf U}_{5,{\mathcal A}_5}^{d}\right){\bf W}_{5,{\mathcal A}_5}{\bf s}_{5,{\mathcal A}_5}\nonumber \\
            &+\left({\bf H}_{5,b}{\bf U}_{5,{\mathcal A}_6}^{b}+{\bf H}_{5,c}{\bf U}_{5,{\mathcal A}_6}^{c}\right){\bf W}_{5,{\mathcal A}_6}{\bf s}_{5,{\mathcal A}_6} \nonumber \\
            &+\sum\limits_{i',j',k:{\bf G}_{5,{\mathcal A}_k}^{(i',j')}\in {\mathcal J}_5}{\bf G}^{(i',j')}_{5,{\mathcal A}_k}{\bf W}_{(i',j'),{\mathcal A}_k}{\bf s}_{(i',j'),{\mathcal A}_k},
\end{align}
where the former six items are desired signals and the remaining items in the summation are interference. Fig. \ref{signal space} illustrates the signal space of the received signal at user 5.

\begin{remark}\label{dimension design for mu=1/2}
  For each user $(i,j)$, the number of nonzero effective channels in ${\mathcal J}_{(i,j)}$ is $8(\sqrt{|\Psi|}-1)+2(|\Psi|-1)$. Note that the construction of ${\mathcal W}(n)$ should contain all effective channels in ${\mathcal J}\triangleq \bigcup\limits_{(i,j)\in\Psi}{\mathcal J}_{(i,j)}$ so as to achieve interference alignment at every user. Thus, the signal dimension is $M=|{\mathcal W}(n)|=n^{8|\Psi|(\sqrt{|\Psi|}-1)+2|\Psi|(|\Psi|-1)}$. Considering that the total dimension of six desired signals is $6M$ and the dimension of all interference vectors in set ${\mathcal W}(n+1)$ is $(n+1)^{8|\Psi|(\sqrt{|\Psi|}-1)+2|\Psi|(|\Psi|-1)}$, the symbol extension number should be $N=6n^{8|\Psi|(\sqrt{|\Psi|}-1)+2|\Psi|(|\Psi|-1)}+(n+1)^{8|\Psi|(\sqrt{|\Psi|}-1)+2|\Psi|(|\Psi|-1)}$.
\end{remark}

Next, we show that for each user, the received vectors of desired signals are linearly independent from those of interference to ensure the decodability. Taking user 5 as an example, we denote
\begin{align}
{\bf D}_5=\Big[&\left({\bf H}_{5,a}{\bf H}_{2,b}-{\bf H}_{5,b}{\bf H}_{2,a}\right){\bf W}_{5,{\mathcal A}_1} \nonumber \\
               &\left({\bf H}_{5,c}{\bf H}_{8,d}-{\bf H}_{5,d}{\bf H}_{8,c}\right){\bf W}_{5,{\mathcal A}_2} \nonumber \\
               &\left({\bf H}_{5,a}{\bf H}_{4,c}-{\bf H}_{5,c}{\bf H}_{4,a}\right){\bf W}_{5,{\mathcal A}_3} \nonumber \\
               &\left({\bf H}_{5,b}{\bf H}_{6,d}-{\bf H}_{5,d}{\bf H}_{6,b}\right){\bf W}_{5,{\mathcal A}_4} \nonumber \\
               &\left({\bf H}_{5,a}{\bf U}_{5,{\mathcal A}_5}^{a}+{\bf H}_{5,c}{\bf U}_{5,{\mathcal A}_5}^{c}\right){\bf W}_{5,{\mathcal A}_5}  \nonumber \\
               &\left({\bf H}_{5,b}{\bf U}_{5,{\mathcal A}_6}^{b}+{\bf H}_{5,d}{\bf U}_{5,{\mathcal A}_6}^{d}\right){\bf W}_{5,{\mathcal A}_6} \Big],
\end{align}
 and shall prove the full rankness of matrix
\begin{align}\label{Lambda5}
  {\bf \Lambda}_{5}=\left[{\bf D}_5 \quad {\bf W}'\right].
\end{align}
 The proof of the decodability for the entire cellular network is placed in Appendix \ref{appendix_partial_cooperation}.

Finally, since each user can decode $6n^{8|\Psi|(\sqrt{|\Psi|}-1)+2|\Psi|(|\Psi|-1)}$ desired signals over $6n^{8|\Psi|(\sqrt{|\Psi|}-1)+2|\Psi|(|\Psi|-1)}$ $+(n+1)^{8|\Psi|(\sqrt{|\Psi|}-1)+2|\Psi|(|\Psi|-1)}$
symbol extensions, the achievable per-user DoF can be calculated as
\begin{align}
  &d=\frac{6n^{8|\Psi|(\sqrt{|\Psi|}-1)+2|\Psi|(|\Psi|-1)}}{6n^{8|\Psi|(\sqrt{|\Psi|}-1)+2|\Psi|(|\Psi|-1)}\hspace{-1mm}+\hspace{-1mm}(n+1)^{8|\Psi|(\sqrt{|\Psi|}-1)+2|\Psi|(|\Psi|-1)}}.
\end{align}
Taking $n\rightarrow\infty$, we can obtain that the achievable per-user DoF is $d=\frac{6}{7}$ and the proof of corner point $(\mu=1/2,1/d=7/6)$ is completed.

\begin{remark}
It is worth pointing out that the above delivery scheme of the cellular network differs from the cooperative transmission schemes of fully connected networks \cite{Ali2,xu,Ali3,long_journal,CoMP,DoF_CoMP}. In the proposed delivery scheme with partial BS cooperation (also the scheme with full BS cooperation introduced in the next subsection), we not only let the nearby BSs ${\mathcal M}_{(i,j)}$ cooperatively serve user $(i,j)$, but also let other BSs transmit the common messages desired by user $(i,j)$ to increase per-user DoF of the network by interference neutralization and interference alignment.
\end{remark}

\subsection{$\mu=1$ (Full BS Cooperation)}
In this case, all BSs can fully cooperate to serve users. The system thus can be regarded as a partially connected virtual MISO broadcast channel. We denote all the channel coefficients $h_{(i,j),(p,q)}$ as a $|\Psi| \times |\Phi|$ matrix ${\bf H}_{\Psi,\Phi}$, with each row denoting the channel coefficients from all BSs to each user. Since each user can only be connected to the four nearby BSs, there are four nonzero elements in each row of ${\bf H}_{\Psi,\Phi}$. Note that the positions of nonzero elements in each row of ${\bf H}_{\Psi,\Phi}$ are different and all the channel coefficients are independent of each other. Therefore, the matrix ${\bf H}_{\Psi,\Phi}$ has full rank with probability one. A zero-forcing precoding matrix ${\bf H}^{\dag}_{\Psi,\Phi}$ can be used to neutralize all the interference. Thus, we can achieve per-user DoF $d=1$ and the proof of corner point $(\mu=1,1/d=1)$ is completed.

\subsection{Comparison with the scheme in \cite{partially_cellular}}
\begin{figure}[t]
    \begin{centering}
  \includegraphics[scale=0.55]{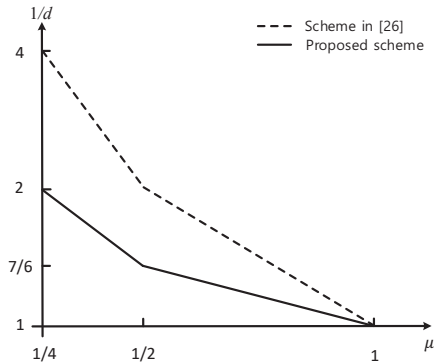}
   \caption{The reciprocal of per-user DoF versus normalized cache size. The solid line is achieved by using the proposed scheme while the dash line is achieved by the scheme in \cite{partially_cellular}.} \label{DoF_versus_mu}
   \end{centering}
  \end{figure}

In this subsection, we compare the proposed scheme with the existing scheme \cite{partially_cellular}\footnote{The assumption of \cite{partially_cellular} is different from ours. In specific, both the achievability and converse in \cite{partially_cellular} are restricted to linear one-shot processing while we allow asymptotic interference alignment and interference neutralization to explore the optimal transmission DoF.} on cache-aided cellular networks. The main idea of our proposed delivery scheme is to utilize the common cached contents among BSs for interference neutralization (at $\mu=\frac{1}{2}$ and $1$), and exploit the distinct cached contents among them to achieve interference alignment gain (at $\mu=\frac{1}{4}$ and $\frac{1}{2}$). However, the transmission scheme in \cite{partially_cellular} is to only activate the BSs caching the same requested file segment at each time and use a zero-forcing precoding to do interference neutralization among different data streams. If applying the scheme\cite{partially_cellular} in our considered network model, we would obtain worse per-user DoF results. In specific, for $\mu=\frac{1}{4}$, the BSs in ${\mathcal B}_1$, ${\mathcal B}_2$, ${\mathcal B}_3$ and ${\mathcal B}_4$ are activated alternately and the per-user DoF $\tilde{d}=\frac{1}{4}$ can be obtained; for $\mu=\frac{1}{2}$, the BSs in ${\mathcal B}_1\cup{\mathcal B}_2$, ${\mathcal B}_1\cup{\mathcal B}_3$, ${\mathcal B}_1\cup{\mathcal B}_4$, ${\mathcal B}_2\cup{\mathcal B}_3$, ${\mathcal B}_2\cup{\mathcal B}_4$ and ${\mathcal B}_3\cup{\mathcal B}_4$ are activated alternately and the per-user DoF $\tilde{d}=\frac{1}{2}$ can be obtained; for $\mu=1$, all the BSs in $\Phi$ are activated and the per-user DoF $\tilde{d}=1$ can be obtained. By using memory sharing, the per-user DoF achieved by the scheme in \cite{partially_cellular} satisfies
  \begin{align}
    1/\tilde{d}=\left\{\begin{array}{ll}
6-8\mu, & \mu\in\left[\frac{1}{4},\frac{1}{2}\right), \\
 3-2\mu, & \mu\in\left[\frac{1}{2},1\right]. \\
\end{array}\right.
  \end{align}
  Fig. \ref{DoF_versus_mu} shows the curves of the reciprocal of per-user DoF versus normalized cache size $\mu$. It can be seen that our proposed scheme is superior in the whole cache size region.

\section{Converse}
In this section, we present an upper bound of the per-user DoF for the considered network under the given cache placement scheme in Section \ref{cache_model}. Based on this upper bound, we show that our achievable per-user DoF in Theorem \ref{theorem1} is optimal in certain cache size region and within a constant additive gap to the optimum in the remaining cache size region. Note that since each user is equipped with one antenna, the cut-set upper bound of the per-user DoF is $d\leq 1$ for all possible caching and delivery strategies. This means that our achievable per-user DoF is within a multiplicative factor of 2 to the optimum, which, however is too loose to demonstrate the effectiveness of our proposed delivery scheme. To obtain the new upper bound tighter than the cut-set bound, we first derive an outer bound on the DoF region of the considered network for any given uncoded prefetching scheme. This outer bound is not restricted to the cellular network with grid topology and can be applied to networks with other general topologies. We then obtain an upper bound of the per-user DoF and show the order-optimality of the proposed delivery scheme with a constant additive gap.

\subsection{Outer Bound on DoF Region}
 Consider a cellular network for any given uncoded prefetching scheme with BS cache groups $\Theta=\{{\mathcal A}_k\}$. We define $R_{(i,j),{\mathcal A}_k}(\mu,P)\triangleq\frac{|W_{r_{(i,j)}, {\mathcal A}_k}|}{T}$ as the rate of message $W_{r_{(i,j)}, {\mathcal A}_k}$ which is transmitted from BS cache group ${\mathcal A}_k$ to user $(i,j)$. Define capacity region ${\mathcal C}(\mu,P)$ as the set of all achievable rate tuples ${\mathfrak R}=\{R_{(i,j),{\mathcal A}_k}(\mu,P)\}_{(i,j)\in\Psi,{\mathcal A}_k\in\Theta}$. The DoF region of the cache-aided cellular network is defined as
\begin{align}\label{DoF_region}
  {\mathcal D}\triangleq&\Bigg\{{\bf d}=\{d_{(i,j),{\mathcal A}_k}\}_{(i,j)\in\Psi,{\mathcal A}_k\in\Theta}\in {\mathbb R}^{|\Psi||\Theta|}_{+}:\forall \,  \{\omega_{(i,j),{\mathcal A}_k}\}_{(i,j)\in\Psi,{\mathcal A}_k\in\Theta}\in {\mathbb R}^{|\Psi||\Theta|}_{+}, \nonumber \\
  &\sum\limits_{(i,j)\in\Psi,{\mathcal A}_k\in\Theta} \omega_{(i,j),{\mathcal A}_k}d_{(i,j),{\mathcal A}_k}\leq \mathop{\lim \sup}\limits_{P\rightarrow \infty} \Bigg[\sup_{{\mathfrak R}\in{\mathcal C}}\Bigg[\sum\limits_{(i,j)\in\Psi,{\mathcal A}_k\in\Theta}\omega_{(i,j),{\mathcal A}_k}R_{(i,j),{\mathcal A}_k}(\mu,P)\Bigg]\frac{1}{\log P}\Bigg] \Bigg\}.
\end{align}

\begin{theorem}\label{DoF_upper_theorem}
Consider a cache-aided cellular network with general topology. For each arbitrary user set ${\mathcal R}$, we define a corresponding BS set $\mathcal T$ which satisfies the following two conditions:
  \begin{enumerate}
    \item $|\mathcal T|=|\mathcal R|$.
    \item The matrix ${\bf H}_{\mathcal R,\mathcal T}$, with each row denoting the channel coefficients from all the BSs in $\mathcal T$ to each user in $\mathcal R$, has full rank almost surely. Note that the channel coefficient from a BS to a user will be zero if there is no connection between them.
  \end{enumerate}
  Denote all the pairs $(\mathcal R,\mathcal T)$ as a set $\mathcal S$. Then, an outer bound on the DoF region of the considered network for any given uncoded prefetching scheme with BS cache groups $\Theta=\{{\mathcal A}_k\}$ is given by
  \begin{align}\label{inequality_dof_upper}
    {\mathcal D}_{\text{out}}=&\bigg\{{\bf d}\in {\mathbb R}^{|\Psi||\Theta|}_{+}:\sum\limits_{(i,j)\in {\mathcal R}}\sum\limits_{{{\mathcal A}_k} \in \Theta}d_{(i,j),{{\mathcal A}_k}}+\sum\limits_{(i,j)\in \bar{\mathcal R}}\sum\limits_{{{\mathcal A}_k} \subseteq {\mathcal T}, {{\mathcal A}_k} \in \Theta}d_{(i,j),{{\mathcal A}_k}}\leq |\mathcal R|, \forall (\mathcal R,\mathcal T)\in{\mathcal S}\bigg\},
  \end{align}
 where $\bar{\mathcal R}$ denotes the set of the remaining users connected to the BSs in $\mathcal T$ except the users in $\mathcal R$.
\end{theorem}
\begin{IEEEproof}
Please see Appendix \ref{proof DoF upper theorem}.
\end{IEEEproof}
  \begin{figure}[t]
    \begin{centering}
  \includegraphics[scale=0.48]{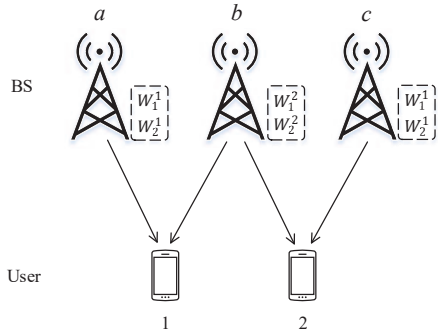}
   \caption{An example network where user 1 is connected to BSs $a$ and $b$, and user 2 is connected to BSs $b$ and $c$. Assume that there are two files $W_1$ and $W_2$ in the library. The prefetching scheme is that each file $W_i$, for $i\in[2]$, is equally split into two disjoint parts $\{W^1_i, W^2_i\}$, with $W^1_i$ cached at BSs $a$ and $c$, and $W^2_i$ cached at BS $b$. } \label{converse example}
   \end{centering}
    \end{figure}

To illustrate how to apply Theorem \ref{DoF_upper_theorem}, we consider an example network with a given uncoded prefetching scheme as shown in Fig. \ref{converse example}. Here, the set $\Theta$ contains two BS cache groups, i.e., ${\mathcal A}_1=\{a,c\}$ and ${\mathcal A}_2=\{b\}$. The pairs $(\mathcal R,\mathcal T)$ satisfying the two conditions in Theorem \ref{DoF_upper_theorem} are
\begin{align}
  {\mathcal S}=\Big\{&\big(\{1\},\{a\}\big),\big(\{1\},\{b\}\big),\big(\{2\},\{b\}\big),\big(\{2\},\{c\}\big),\nonumber \\
  &\big(\{1,2\},\{a,b\}\big),\big(\{1,2\},\{b,c\}\big),\big(\{1,2\},\{a,c\}\big)\Big\}.
\end{align}
By substituting $(\mathcal R,\mathcal T)=\big(\{1\},\{a\}\big),\big(\{1\},\{b\}\big),\big(\{2\},\{b\}\big),\big(\{2\},\{c\}\big)$ into the inequality in \eqref{inequality_dof_upper}, we obtain
\begin{subequations}
\begin{align}
  &d_{1,{\mathcal A}_1}+d_{1,{\mathcal A}_2}\leq1,\label{inequality 1}\\
  &d_{1,{\mathcal A}_1}+d_{1,{\mathcal A}_2}+d_{2,{\mathcal A}_2}\leq1,\label{inequality 2}\\
  &d_{2,{\mathcal A}_1}+d_{2,{\mathcal A}_2}+d_{1,{\mathcal A}_2}\leq1,\label{inequality 3}\\
  &d_{2,{\mathcal A}_1}+d_{2,{\mathcal A}_2}\leq1,\label{inequality 4}
\end{align}
\end{subequations}
respectively. It is obvious that the inequalities \eqref{inequality 1} and \eqref{inequality 4} are redundant. By substituting $\big(\{1,2\},\{a,b\}\big)$, $\big(\{1,2\},\{b,c\}\big),\big(\{1,2\},\{a,c\}\big)$ into the inequality \eqref{inequality_dof_upper}, we obtain
\begin{align}\label{inequality 5}
 d_{1,{\mathcal A}_1}+d_{1,{\mathcal A}_2}+d_{2,{\mathcal A}_1}+d_{2,{\mathcal A}_2}\leq2.
\end{align}
 Via adding \eqref{inequality 2} and \eqref{inequality 3}, we can obtain a tighter bound than \eqref{inequality 5} and thus \eqref{inequality 5} is redundant. The outer bound ${\mathcal D}_{\text{out}}$ of the network in Fig. \ref{converse example} for the cache placement scheme with BS cache groups $\Theta=\{{\mathcal A}_1,{\mathcal A}_2\}$ is given by
  \begin{equation}
    {\mathcal D}_{\text{out}}=\left\{{\bf d}\in {\mathbb R}^4_{+}\left|\begin{array}{c}
                                                                         d_{1,{\mathcal A}_1}+d_{1,{\mathcal A}_2}+d_{2,{\mathcal A}_2}\leq1 \\
                                                                         d_{2,{\mathcal A}_1}+d_{2,{\mathcal A}_2}+d_{1,{\mathcal A}_2}\leq1
                                                                       \end{array}
    \right.\right\}.
  \end{equation}

\subsection{Upper Bound on Per-User DoF}
 Next, we use Theorem \ref{DoF_upper_theorem} to derive an upper bound on the per-user DoF of the considered cellular network for the given cache placement scheme in Section \ref{cache_model}. Recalling that $\Theta$ is the set of all BS cache groups $\{{\mathcal A}_k\}$, the first item $\sum\limits_{(i,j)\in {\mathcal R}}\sum\limits_{{{\mathcal A}_k} \in \Theta}d_{(i,j),{{\mathcal A}_k}}$ in \eqref{inequality_dof_upper} is equal to $|\mathcal R|\cdot d$ with $d$ being the per-user DoF. Since each user requests the same number of subfiles from the BS cache groups in the adopted caching scheme, the second item $\sum\limits_{(i,j)\in \bar{\mathcal R}}\sum\limits_{{{\mathcal A}_k} \subseteq {\mathcal T}, {{\mathcal A}_k} \in \Theta}d_{(i,j),{{\mathcal A}_k}}$ in \eqref{inequality_dof_upper} can be expressed as $|\bar{\mathcal R}|\cdot\sum\limits_{{{\mathcal A}_k} \subseteq {\mathcal T}, {{\mathcal A}_k} \in \Theta}d_{(i,j),{{\mathcal A}_k}}$. By dividing both sides of the inequality \eqref{inequality_dof_upper} with $|\mathcal R|$, we have
 \begin{align}\label{inequality_temp}
  d+\frac{|\bar{\mathcal R}|}{|\mathcal R|}\sum\limits_{{{\mathcal A}_k} \subseteq {\mathcal T}, {{\mathcal A}_k} \in \Theta}d_{(i,j),{{\mathcal A}_k}}\leq 1, \quad \forall (\mathcal R,\mathcal T)\in{\mathcal S}.
 \end{align}
 By defining a function
 \begin{align}
  \lambda(\mathcal R,\mathcal T)\triangleq\frac{1}{d}\cdot \frac{|\bar{\mathcal R}|}{{|\mathcal R|}}\sum\limits_{{{\mathcal A}_k} \subseteq {\mathcal T}, {{\mathcal A}_k} \in \Theta}d_{(i,j),{{\mathcal A}_k}},
 \end{align}
 the inequality \eqref{inequality_temp} can be rewritten as $d+\lambda(\mathcal R,\mathcal T)\cdot d=d(1+\lambda(\mathcal R,\mathcal T))\leq 1$. To obtain the tightest upper bound of per-user DoF $d$, we should find the maximum value of $\lambda(\mathcal R,\mathcal T)$, i.e.
    \begin{align}
  \mathcal{P}: \quad &\max_{(\mathcal R,\mathcal T)\in{\mathcal S}} \ \lambda(\mathcal R,\mathcal T)
 \end{align}
 Based on the specific file splitting and placement strategy in Section \ref{cache_model}, we can determine the optimal $\mathcal T^*$ as shown in the following lemma.
\begin{lemma}\label{bs_t}
  For the considered cache-aided cellular network with the given prefetching scheme in Section \ref{cache_model}, the optimal $\mathcal T^*$ in problem $\mathcal P$ satisfies one of the following forms:
  \begin{enumerate}
   \item ${\mathcal T}^*\in\{{\mathcal B}_1,{\mathcal B}_2,{\mathcal B}_3,{\mathcal B}_4\}$;
   \item ${\mathcal T}^*\in\{{\mathcal B}_1\cup{\mathcal B}_2,{\mathcal B}_1\cup{\mathcal B}_3,{\mathcal B}_1\cup{\mathcal B}_4,{\mathcal B}_2\cup{\mathcal B}_3,{\mathcal B}_2\cup{\mathcal B}_4,,{\mathcal B}_3\cup{\mathcal B}_4\}$;
   \item ${\mathcal T}^*\in\{\Phi\backslash{\mathcal B}_1,\Phi\backslash{\mathcal B}_2,\Phi\backslash{\mathcal B}_3,\Phi\backslash{\mathcal B}_4\}$;
   \item ${\mathcal T}^*=\Phi$.
  \end{enumerate}
\end{lemma}
\begin{IEEEproof}
  Please see Appendix \ref{proof_bs_t}.
\end{IEEEproof}

\begin{figure}[t]
    \begin{centering}
  \includegraphics[scale=0.55]{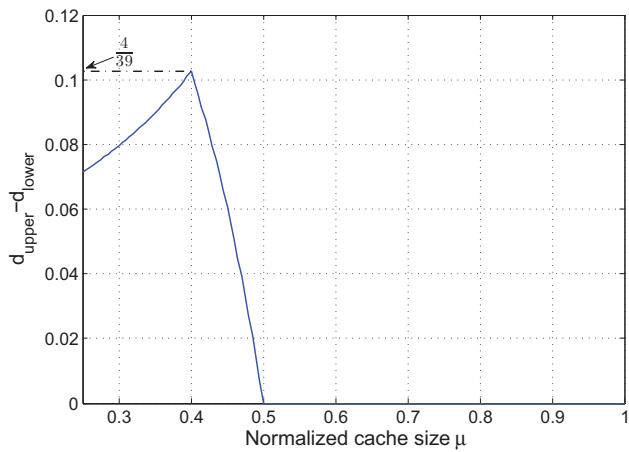}
   \caption{The curve of the additive gap versus normalized cache size $\mu$. } \label{gap}
   \end{centering}
    \end{figure}

Since each ${\mathcal T}^*$ in Lemma \ref{bs_t} is connected to all users $\Psi$, the corresponding ${\mathcal R}^*$ satisfies ${\mathcal R}^*\cup\bar{{\mathcal R}}^*=\Psi$. For ${\mathcal T}^*$ with size $|\mathcal T^*|= \frac{|\Phi|}{4}$, $\frac{|\Phi|}{2}$,
 $\frac{3|\Phi|}{4}$ and $|\Phi|$, the value $\frac{|\bar{{\mathcal R}}^*|}{|\mathcal R^*|}$ is determined and given by $3$, $1$, $\frac{1}{3}$ and $0$, respectively. In other words, the optimal value $\lambda^*$ only depends on set ${\mathcal T^*}$ and the cardinality $|{\mathcal R}^*|$ (which is equal to $|{\mathcal T}^*|$), but not on the specific users in ${\mathcal R}^*$. One optimal form of ${\mathcal R}^*$ is (1) ${\mathcal R}^*={\mathcal U}_1$ for $|\mathcal T^*|=\frac{|\Phi|}{4}$, (2) ${\mathcal R}^*={\mathcal U}_1\cup{\mathcal U}_2$ for $|\mathcal T^*|=\frac{|\Phi|}{2}$, (3) ${\mathcal R}^*={\mathcal U}_1\cup{\mathcal U}_2\cup{\mathcal U}_3$ for $|\mathcal T^*|=\frac{3|\Phi|}{4}$ and (4) ${\mathcal R}^*=\Phi$ for $|{\mathcal T}^*|=|\Phi|$,
where $\{{\mathcal U}_i\}_{i\in[4]}$ are defined as
 \begin{subequations}\label{user_set}
  \begin{align}
  {\mathcal U}_1&=\left\{(i,j):i=4m+1,i=4n+1,  m,n \in \mathbb{Z}\right\}, \\
  {\mathcal U}_2&=\left\{(i,j):i=4m-1,j=4n+1,  m,n \in \mathbb{Z}\right\}, \\
  {\mathcal U}_3&=\left\{(i,j):i=4m+1,j=4n-1,  m,n \in \mathbb{Z}\right\}, \\
  {\mathcal U}_4&=\left\{(i,i):i=4m-1,j=4n-1,  m,n \in \mathbb{Z}\right\}.
\end{align}
\end{subequations}
By searching for all possible $(\mathcal R^*,\mathcal T^*)$, we can obtain the upper bound on the per-user DoF.
\begin{theorem}\label{theorem3}
For the considered cache-aided cellular network with each BS having a cache of normalized size $\mu$ and the given prefetching scheme in Section \ref{cache_model}, the reciprocal of the per-user DoF $d$ is lower bounded as
\begin{align}
  1/d\geq 1/d_{\text{upper}}\triangleq\left\{\begin{array}{ll}
  1/\min\left\{\frac{2}{5-6\mu},\frac{6}{11-8\mu}\right\}, & \mu\in\left[\frac{1}{4},\frac{1}{2}\right), \\
  \frac{4}{3}-\frac{1}{3}\mu, & \mu\in\left[\frac{1}{2},1\right]. \\
\end{array}\right.
\end{align}
\end{theorem}
\begin{IEEEproof}
Please see Appendix \ref{proof_theorem3}.
\end{IEEEproof}
\begin{remark}
  Comparing Theorem \ref{theorem1} with Theorem \ref{theorem3}, we can see that the upper and lower bounds of the per-user DoF coincide and thus the proposed delivery scheme is optimal for cache size $\mu\in\left[\frac{1}{2},1\right]$, and their additive gap for cache size $\mu\in\left[\frac{1}{4},\frac{1}{2}\right)$ is given by
  \begin{align}
    d_{\text{upper}}-d_{\text{lower}}=\min\left\{\frac{2}{5-6\mu},\frac{6}{11-8\mu}\right\}-\frac{6}{17-20\mu}\leq\frac{4}{39}.
  \end{align}
  Fig. \ref{gap} shows the exact value of the additive gap between the upper and lower bounds of the per-user DoF at different cache size $\mu$.
\end{remark}

\section{Conclusion}
In this work, we characterized the per-user DoF of the cache-aided cellular network where the locations of BSs are modeled as a grid form and users within a grid cell can only communicate with four nearby BSs. By adopting a file splitting and placement scheme tailored for the network with partial connectivity, there exist different levels of BS cooperation in the delivery. We proposed the transmission schemes with no BS cooperation, partial BS cooperation and full BS cooperation, respectively, for different cache sizes. The achievable DoF results reveal that the reciprocal of per-user DoF for the cache-aided cellular network decreases piecewise linearly with the cache size and the gain of BS caching is more significant for the small cache region. Besides, we showed that under the given cache placement scheme, the obtained per-user DoF of the cache-aided cellular network is optimal when $\mu\in\left[\frac{1}{2},1\right]$, and within an additive gap of $\frac{4}{39}$ to the optimum when $\mu\in\left[\frac{1}{4},\frac{1}{2}\right)$.

\begin{appendices}

\section{Proof of Point $(\mu=1/4,1/d=2)$}\label{appendix_no_cooperation}
 The total transmission takes $4n^{3|\Psi|}+4(n+1)^{3|\Psi|}$ symbol extensions. Each BS $(p,q)$ employs a random Gaussian coding scheme to encode the four subfiles $W_{r_{(p-1,q+1)},{\mathcal A}_k}$, $W_{r_{(p+1,q+1)},{\mathcal A}_k}$, $W_{r_{(p+1,q-1)},{\mathcal A}_k}$ and $W_{r_{(p-1,q-1)},{\mathcal A}_k}$ requested by its serving users for $(p,q)\in{{\mathcal A}_k}$, into complex signal vectors ${\bf s}_{(p-1,q+1),(p,q)}$, ${\bf s}_{(p+1,q+1),(p,q)}$, ${\bf s}_{(p+1,q-1),(p,q)}$ and ${\bf s}_{(p-1,q-1),(p,q)}$, respectively, each with dimension $M=n^{3|\Psi|}$. The whole transmission contains four phases and each phase takes $N= n^{3|\Psi|}+(n+1)^{3|\Psi|}$ symbol extensions. In each phase, each BS transmits one of the above signals to serve one user as illustrated in Fig. \ref{alternating}. Without loss of generality, we take the first phase as an example.

 The received signal (ignoring noise) of user $(i,j)$ can be represented as
\begin{align}
{\bf y}_{(i,j)}&={\bf H}_{(i,j),(i+1,j-1)}{\bf V}_{(i,j),(i+1,j-1)}{\bf s}_{(i,j),(i+1,j-1)\nonumber}\\
&\quad +{\bf H}_{(i,j),(i-1,j-1)}{\bf V}_{(i-2,j),(i-1,j-1)}{\bf s}_{(i-2,j),(i-1,j-1)}\nonumber\\
&\quad+{\bf H}_{(i,j),(i-1,j+1)}{\bf V}_{(i-2,j+2),(i-1,j+1)}{\bf s}_{(i-2,j+2),(i-1,j+1)}\nonumber\\
&\quad+{\bf H}_{(i,j),(i+1,j+1)}{\bf V}_{(i,j+2),(i+1,j+1)}{\bf s}_{(i,j+2),(i+1,j+1)},
\end{align}
where the first item is the desired signal and the remaining three items are interference signals. For each user $(i,j)$, we aim to align the directions of three interference signals into a common space and the interference alignment conditions are given by
\begin{align}
  &\textrm{span}\left({\bf H}_{(i,j),(i-1,j-1)}{\bf V}_{(i-2,j),(i-1,j-1)}\right)\subset  \textrm{span}\left({\bf V}'\right), \nonumber\\
  &\textrm{span}\left({\bf H}_{(i,j),(i-1,j+1)}{\bf V}_{(i-2,j+2),(i-1,j+1)}\right)\subset \textrm{span}\left({\bf V}'\right), \nonumber \\
  &\textrm{span}\left({\bf H}_{(i,j),(i+1,j+1)}{\bf V}_{(i,j+2),(i+1,j+1)}\right)\subset
\textrm{span}\left({\bf V}'\right),\quad \forall (i,j)\in\Psi.
\end{align}
We collect the channel matrices experienced by interference for all users as
\begin{align}
{\mathcal I}=\{{\bf H}&_{(i,j),(p,q)}:(p,q)\in{\mathcal M}_{(i,j)}\setminus\{(i+1,j-1)\}, (i,j)\in\Psi \},
\end{align}
and construct
\begin{align}
{\mathcal V}(n)=\Bigg\{\Bigg(\prod\limits_{{\bf H}_{(i,j),(p,q)}\in{\mathcal I}} \Big({\bf H}_{(i,j),(p,q)}&\Big)^{\iota_{(i,j),(p,q)}}{\bf 1}_{N}\Bigg):\iota_{(i,j),(p,q)}\in[n]\Bigg\}.
\end{align}
The column vectors in ${\mathcal V}(n)$ can occur in any order to form each precoding matrix ${\bf V}_{(i,j),(p,q)}$. Specifically, the $m$-th column of ${\bf V}_{(i,j),(p,q)}$, ${\bf v}^m_{(i,j),(p,q)}$ can be expressed as
\begin{align}
  {\bf v}^m_{(i,j),(p,q)}=\prod\limits_{{\bf H}_{(i',j'),(p',q')}\in{\mathcal I}} \Big({\bf H}_{(i',j'),(p',q')}\Big)^{\iota_{(i',j'),(p',q')}^{m}}{\bf 1}_{N},
\end{align}
where the subscripts of the exponent $\iota_{(i',j'),(p',q')}^{m}$ are associated
with corresponding base ${\bf H}_{(i',j'),(p',q')}$, and the superscript of $\iota_{(i',j'),(p',q')}^{m}$, $m$ indicates the column index of ${\bf v}^m_{(i,j),(p,q)}$ in ${\bf V}_{(i,j),(p,q)}$. The received signal at user $(i,j)$ can be rewritten as 
\begin{align}\label{v_design_signal}
  {\bf y}_{(i,j)}=&\underbrace{{\bf H}_{(i,j),(i+1,j-1)}\sum\limits_{m=1}^{M}\prod\limits_{{\bf H}_{(i',j'),(p',q')}\in{\mathcal I}} \Big({\bf H}_{(i',j'),(p',q')}\Big)^{\iota_{(i',j'),(p',q')}^{m}}{\bf 1}_{N}s_{(i+1,j-1)}^{m}}_{\text{desired signal}}\nonumber \\
  &+\underbrace{\sum\limits_{(p,q)\in{\mathcal M}_{(i,j)}\setminus\{(i+1,j-1)\}}{\bf H}_{(i,j),(p,q)}\sum\limits_{m=1}^{M} \prod\limits_{{\bf H}_{(i',j'),(p',q')}\in{\mathcal I}} \Big({\bf H}_{(i',j'),(p',q')}\Big)^{\iota_{(i',j'),(p',q')}^{m}}{\bf 1}_{N}s_{(p,q)}^{m}}_{\text{interference}}.
\end{align}
 Note that in \eqref{v_design_signal}, all the channel matrices experienced by interference, i.e., $\{{\bf H}_{(i,j),(p,q)}\}$ with $(p,q)\in{\mathcal M}_{(i,j)}\setminus\{(i+1,j-1)\}$, belong to the set $\mathcal I$. Thus, the received precoding vectors of interference are aligned into the set ${\mathcal V}(n+1)$.

\begin{table*}
\newcommand{\tabincell}[2]{\begin{tabular}{@{}#1@{}}#2\end{tabular}}  %������
\centering
\caption{The design of ${\bf U}_{(i,j),{\mathcal A}_k}^{(p,q)}$ for $(i,j)\in{\mathcal U}_1$.}\label{table2}
\begin{tabular}{|c|c|c|}
\hline
 BS Cache Group & BS Coordinate & Design Method \\
\hline
\multirow{3}*{${\mathcal A}_1$}  &  ${\mathcal A}_1\setminus\{(p,q):q=j-1\}$  & ${\bf U}_{(i,j),{\mathcal A}_1}^{(p,q)}={\bf 0}$ \\
\cline{2-3}
~  &  $\{(p,q):p\in\{i\pm1\},q=j-1\}$ & ${\bf U}_{(i,j),{{\mathcal A}_1}}^{(i-1,j-1)}={\bf H}_{(i,j-2),(i+1,j-1)}, {\bf U}_{(i,j),{{\mathcal A}_1}}^{(i+1,j-1)}=-{\bf H}_{(i,j-2),(i-1,j-1)}$\\
\cline{2-3}
~  &  $\{(p,q):p\notin\{i\pm1\},q=j-1\}$ & Each diagonal element is i.i.d. from a continuous distribution \\
\hline
\multirow{3}*{${\mathcal A}_2$} & ${\mathcal A}_2\setminus\{(p,q):q=j+1\}$  & ${\bf U}_{(i,j),{\mathcal A}_2}^{(p,q)}={\bf 0}$ \\
\cline{2-3}
~  &  $\{(p,q):p\in\{i\pm1\},q=j+1\}$ & ${\bf U}_{(i,j),{{\mathcal A}_2}}^{(i-1,j+1)}={\bf H}_{(i,j+2),(i+1,j+1)},{\bf U}_{(i,j),{{\mathcal A}_2}}^{(i+1,j+1)}=-{\bf H}_{(i,j+2),(i-1,j+1)}$\\
\cline{2-3}
~  &  $\{(p,q):p\notin\{i\pm1\},q=j+1\}$ & Each diagonal element is i.i.d. from a continuous distribution \\
\hline
\multirow{3}*{${\mathcal A}_3$}  &  ${\mathcal A}_3\setminus\{(p,q):p=i-1\}$  & ${\bf U}_{(i,j),{\mathcal A}_3}^{(p,q)}={\bf 0}$ \\
\cline{2-3}
~  &  $\{(p,q):p=i-1,q\in\{j\pm1\}\}$ & ${\bf U}_{(i,j),{{\mathcal A}_3}}^{(i-1,j-1)}={\bf H}_{(i-2,j),(i-1,j+1)},{\bf U}_{(i,j),{{\mathcal A}_3}}^{(i-1,j+1)}=-{\bf H}_{(i-2,j),(i-1,j-1)}$\\
\cline{2-3}
~  &  $\{(p,q):p=i-1,q\notin\{j\pm1\}\}$ & Each diagonal element is i.i.d. from a continuous distribution \\
\hline
\multirow{3}*{${\mathcal A}_4$}  &  ${\mathcal A}_4\setminus\{(p,q):p=i+1\}$  & ${\bf U}_{(i,j),{\mathcal A}_4}^{(p,q)}={\bf 0}$ \\
\cline{2-3}
~  &  $\{(p,q):p=i+1,q\in\{j\pm1\}\}$ & ${\bf U}_{(i,j),{{\mathcal A}_4}}^{(i+1,j-1)}={\bf H}_{(i+2,j),(i+1,j+1)},{\bf U}_{(i,j),{{\mathcal A}_4}}^{(i+1,j+1)}=-{\bf H}_{(i+2,j),(i+1,j-1)}$\\
\cline{2-3}
~  &  $\{(p,q):p=i+1,q\notin\{j\pm1\}\}$ & Each diagonal element is i.i.d. from a continuous distribution \\
\hline
${\mathcal A}_5,{\mathcal A}_6$ & $(p,q)\in \Phi\cap{{\mathcal A}_k}$ & Each diagonal element is i.i.d. from a continuous distribution \\
\hline
\end{tabular}
\end{table*}

Then, we need to prove the decodability of desired signals. For user $(i,j)$, we denote all the received precoding vectors of desired signals as
\begin{align}
 &{\bf D}_{(i,j)}={\bf H}_{(i,j),(i+1,j-1)}{\bf V}_{(i,j),(i+1,j-1)}.
\end{align}
We should prove that the matrix
\begin{align}
  {\bf \Lambda}_{(i,j)}=\left[{\bf D}_{(i,j)}\quad {\bf V}'\right]
\end{align}
has full rank almost surely. Similar to Section \ref{no_cooperation}, we need to prove that the elements in the same row of ${\bf \Lambda}_{(i,j)}$ are different monomials. Considering an arbitrary $\eta$-th row of ${\bf \Lambda}_{(i,j)}$, (i) the products in ${\bf D}_{(i,j)}$ are different from each other because each ${\bf v}^m_{(i,j),(i+1,j-1)}$ is a unique vector in set ${\mathcal V}(n)$; (ii) since ${\bf H}_{(i,j),(i+1,j-1)}\notin{\mathcal I}$, the items in ${\bf D}_{(i,j)}$ and ${\bf V}'$ differ in channel factor $h_{(i,j),(i+1,j-1)}(\eta)$. Therefore, the full rankness of matrix ${\bf \Lambda}_{(i,j)}$ is proved. The per-user DoF can be calculated as $\frac{n^{|3\Psi|}}{n^{3|\Psi|}+(n+1)^{3|\Psi|}}$. Taking $n\rightarrow\infty$, the achievable per-user DoF $d=\frac{1}{2}$ is obtained.

\section{Proof of Point $(\mu=1/2,1/d=7/6)$}\label{appendix_partial_cooperation}
\subsection{Design of $\{{\bf U}^{(p,q)}_{(i,j),{\mathcal A}_k}\}$}
Note that the relative positions of BS cache groups $\{{\mathcal A}_k\}_{k\in[6]}$ for the users within each of $\{{\mathcal U}_r\}_{r\in[4]}$ (defined in eq. \eqref{user_set}) are the same. Without loss of generality, we elaborate the design of ${\bf U}_{(i,j),{\mathcal A}_k}^{(p,q)}$ for $(i,j)\in{\mathcal U}_1$ (also ${\bf W}_{(i,j),{\mathcal A}_k}$ in next subsection), and the design for users in other ${\mathcal U}_r$ can be obtained by alternating BS cache group indexes. The design of each ${\bf U}_{(i,j),{\mathcal A}_k}^{(p,q)}$ depends on the BS cache group ${\mathcal A}_k$ and BS coordinate $(p,q)$, which is summarized in Table \ref{table2}.

\subsection{Design of $\{{\bf W}_{(i,j),{\mathcal A}_k}\}$}
 In the general cellular network, for user $(i,j)$, the effective channels of interference for each ${\mathcal A}_k$ are given by
\begin{subequations}
  \begin{align}
 &{\mathcal J}_{(i,j),{\mathcal A}_1}=\Big\{{\bf G}_{(i,j),{\mathcal A}_1}^{(i',j')}:i'\neq i,j'\in\{j-2,j\} \Big\},\\
  &{\mathcal J}_{(i,j),{\mathcal A}_2}=\Big\{{\bf G}_{(i,j),{\mathcal A}_2}^{(i',j')}:i'\neq i,j'\in\{j,j+2\} \Big\},\\
  &{\mathcal J}_{(i,j),{\mathcal A}_3}=\Big\{{\bf G}_{(i,j),{\mathcal A}_3}^{(i',j')}:i'\in\{i-2,i\}, j'\neq j \Big\},\\
  &{\mathcal J}_{(i,j),{\mathcal A}_4}=\Big\{{\bf G}_{(i,j),{\mathcal A}_4}^{(i',j')}:i'\in\{i,i+2\}, j'\neq j \Big\},\\
  &{\mathcal J}_{(i,j),{\mathcal A}_5}=\Big\{{\bf G}_{(i,j),{\mathcal A}_5}^{(i',j')}: (i',j')\neq(i,j)\Big\},\\
  &{\mathcal J}_{(i,j),{\mathcal A}_6}=\Big\{{\bf G}_{(i,j),{\mathcal A}_6}^{(i',j')}: (i',j')\neq(i,j)\Big\}.
\end{align}
\end{subequations}
Recall that the cellular network is assumed as a square grid with the number of users in each row (or column) being $\sqrt{|\Psi|}$. Thus, we can obtain $|{\mathcal J}_{(i,j),{\mathcal A}_k}|=2(\sqrt{|\Psi|}-1)$ with $k\in[4]$, and $|{\mathcal J}_{(i,j),{\mathcal A}_k}|=|\Psi|-1$ with $k=5,6$.

We define ${\mathcal J}\triangleq \bigcup\limits_{(i,j)\in\Psi}{\mathcal J}_{(i,j)}=\bigcup\limits_{(i,j)\in\Psi} \{{\mathcal J}_{(i,j),{\mathcal A}_k}\}_{k\in[6]}$. For each user $(i,j)$, we aim to align the directions of all the interference signals into a common space and the interference alignment conditions are expressed as
\begin{align}
 \textrm{span}\left({\bf G}^{(i',j')}_{(i,j),{\mathcal A}_k}{\bf W}_{(i',j'),{\mathcal A}_k}\right)\subset\textrm{span}\left({\bf W}'\right),\quad \forall i',j',k\  \text{satisfy} \ {\bf G}^{(i',j')}_{(i,j),{\mathcal A}_k}\in{\mathcal J}_{(i,j)}.
\end{align}
Towards this end, we construct
\begin{align}
  &{\mathcal W}(n)=\left\{\left(\prod\limits_{\begin{subarray}{c}
   (i,j),(i',j'),k: \\
   {\bf G}_{(i,j),{\mathcal A}_k}^{(i',j')}\in {\mathcal J}
  \end{subarray}
   }\Big({\bf G}_{(i,j),{\mathcal A}_k}^{(i',j')}\Big)^{\iota_{(i,j),{\mathcal A}_k}^{(i',j')}}{\bf 1}_{N}\right):\iota_{(i,j),{\mathcal A}_k}^{(i',j')}\in[n]\right\}.
\end{align}
The column vectors in set ${\mathcal W}(n)$ may occur in any order to form each of matrices $\{{\bf W}_{(i,j),{\mathcal A}_k}\}$.

\subsection{Proof of Decodability}
Before the proof, we introduce the polynomial functions for matrices ${\bf H}_{(i,j),(p,q)}$, ${\bf U}^{(p,q)}_{(i,j),{\mathcal A}_k}$, and ${\bf G}^{(i',j')}_{(i,j),{\mathcal A}_k}$. In specific, the diagonal elements of ${\bf H}_{(i,j),(p,q)}$ can be regarded as the realizations (instances) of channel coefficient $h_{(i,j),(p,q)}$. Considering the specific forms of ${\bf U}_{(i,j),{\mathcal A}_k}^{(p,q)}$ in Table \ref{table2}, the diagonal elements of ${\bf U}_{(i,j),{\mathcal A}_k}^{(p,q)}$ can be viewed as the realizations of a channel coefficient when it is designed as a function of channel matrices, or the realizations of a random variable $u_{(i,j),{\mathcal A}_k}^{(p,q)}$ when its elements are i.i.d. from a continuous distribution, with $u_{(i,j),{\mathcal A}_k}^{(p,q)}$ obeying the same distribution. Similarly, we denote the polynomial function of the diagonal elements in ${\bf G}^{(i',j')}_{(i,j),{\mathcal A}_k}$ as
\begin{align}
  g^{(i',j')}_{(i,j),{\mathcal A}_k}=h_{(i,j),(p,q)}u_{(i',j'),{\mathcal A}_k}^{(p,q)}+h_{(i,j),(p', q')}u_{(i',j'),{\mathcal A}_k}^{(p',q')},
\end{align}
and denote the polynomial function of each column vector in ${\bf W}_{(i,j),{\mathcal A}_k}$ as
\begin{align}\label{form_of_w}
 w_{(i,j),{\mathcal A}_k}=\prod\limits_{\begin{subarray}{c}
   (i',j'),(i'',j''),k: \\
   {\bf G}_{(i',j'),{\mathcal A}_k}^{(i'',j'')}\in {\mathcal J}
  \end{subarray}
} \left(g_{(i',j'),{\mathcal A}_k}^{(i'',j'')}\right)^{\iota_{(i',j'),{\mathcal A}_k}^{(i'',j'')}},
\end{align}
 with $\iota_{(i',j'),{\mathcal A}_k}^{(i'',j'')}\in[n]$. Besides, we denote $w'$ as the polynomial function of each vector in set ${\mathcal W}(n+1)$, which takes the similar form as \eqref{form_of_w} with exponent $\iota_{(i',j'),{\mathcal A}_k}^{(i'',j'')}\in[n+1]$. For convenience, we define
\begin{align}
  \left({\bf g}^r_{(i,j),k}\right)^{{\bm \iota}^r_{(i,j),k}}\triangleq\prod\limits_{(i',j')\in{\mathcal U}_r\setminus\{(i,j)\}}\left(g^{(i',j')}_{(i,j),{\mathcal A}_k}\right)^{\iota^{(i',j')}_{(i,j),{\mathcal A}_k}},
\end{align}
where ${\bf g}^r_{(i,j),k}$ is the polynomial function sequence and ${\bm \iota}^r_{(i,j),k}$ is the exponent sequence. In the following, for brevity, we use index $k$ instead of ${\mathcal A}_k$ for $k\in[6]$, and use index $r$ instead of ${\mathcal U}_r$ for $r\in[4]$ with slight abuse of notation.

Here, we still use the simplified notations in Fig. \ref{simplified} for those BSs and users and show that the decodability for user 5 without loss of generality. According to \cite[Lemma 3]{DoF_CoMP}, we shall prove the linear independence of the following polynomial functions
\begin{align}\label{polynomial_functions}
\begin{array}{ll}
  (h_{5,a}h_{2,b}-h_{5,b}h_{2,a})w_{5,1}, & (h_{5,c}h_{8,d}-h_{5,d}h_{8,c})w_{5,2}, \\
  (h_{5,a}h_{4,c}-h_{5,c}h_{4,a})w_{5,3}, & (h_{5,b}h_{6,d}-h_{5,d}h_{6,b})w_{5,4}, \\
 (h_{5,a}u^a_{5,5}+h_{5,c}u^c_{5,5})w_{5,5}, & (h_{5,b}u^b_{5,6}+h_{5,d}u^d_{5,6})w_{5,6}, \\
  \left\{w'\right\}_{{\bf w}'\in{\mathcal W}(n+1)}, & \
\end{array}
\end{align}
 where each $w_{5,k}$ with $k\in[6]$ is given by
\begin{align}
w_{5,k}=\prod\limits_{\forall (i,j), \forall k,\forall r}\left({\bf g}^r_{(i,j),k}\right)^{{\bm \iota}^r_{(i,j),k}}, \quad {\bm \iota}^r_{(i,j),k}\in[n]^{|\Psi|/4-1},
\end{align}
and each $w'$ is given by
\begin{align}
w'=\prod\limits_{\forall (i,j), \forall k,\forall r}\left({\bf g}^r_{(i,j),k}\right)^{{\bm \iota}^r_{(i,j),k}}, \quad {\bm \iota}^r_{(i,j),k}\in[n+1]^{|\Psi|/4-1}.
\end{align}
Denote the sum of the elements in ${\bm \iota}^r_{(i,j),k}$ as $s({\bm \iota}^r_{(i,j),k})$. The polynomial functions in \eqref{polynomial_functions}, which have different sum exponent $\sum\limits_{(i,j),k,r}s({\bm \iota}^r_{(i,j),k})$, are linearly independent. Thus, we shall consider the worst case where all the $\sum\limits_{(i,j),k,r}s({\bm \iota}^r_{(i,j),k})$ are the same. Due to that each of $\{{\bf g}^r_{(i,j),k}\}$ has at least one different variable $h_{(i,j),(p,q)}$ (or $u_{(i,j),k}^{(p,q)}$) from others, we should prove the linear independence of polynomial functions within each of the following sets
\begin{subequations}
\begin{align}
  &\left\{ (h_{5,a}h_{2,b}-h_{5,b}h_{2,a})\left({\bf g}^1_{5,1}\right)^{{\bm \iota}^1_{5,1}}: s({\bm \iota}^1_{5,1})=E_{5,1}^1-1\right\}\cup \left\{\left({\bf g}^1_{5,1}\right)^{{\bm \iota}^1_{5,1}}: s({\bm \iota}^1_{5,1})=E_{5,1}^1\right\},\label{g_1}\\
  &\left\{ (h_{5,c}h_{8,d}-h_{5,d}h_{8,c})\left({\bf g}^1_{5,2}\right)^{{\bm \iota}^1_{5,2}}: s({\bm \iota}^1_{5,2})=E_{5,2}^1-1\right\}\cup \left\{\left({\bf g}^1_{5,2}\right)^{{\bm \iota}^1_{5,2}}: s({\bm \iota}^1_{5,2})=E_{5,2}^1\right\}, \label{g_2}\\
  &\left\{ (h_{5,a}h_{4,c}-h_{5,c}h_{4,a})\left({\bf g}^1_{5,3}\right)^{{\bm \iota}^1_{5,3}}: s({\bm \iota}^1_{5,3})=E_{5,3}^1-1\right\}\cup \left\{\left({\bf g}^1_{5,3}\right)^{{\bm \iota}^1_{5,3}}: s({\bm \iota}^1_{5,3})=E_{5,3}^1\right\},\label{g_3}\\
  &\left\{ (h_{5,b}h_{6,d}-h_{5,d}h_{6,b})\left({\bf g}^1_{5,4}\right)^{{\bm \iota}^1_{5,4}}: s({\bm \iota}^1_{5,4})=E_{5,4}^1-1\right\}\cup \left\{\left({\bf g}^1_{5,4}\right)^{{\bm \iota}^1_{5,4}}: s({\bm \iota}^1_{5,4})=E_{5,4}^1\right\},\label{g_4}\\
  &\left\{ (h_{5,a}u^a_{5,5}+h_{5,c}u^c_{5,5})\left({\bf g}^1_{5,5}\right)^{{\bm \iota}^1_{5,5}}: s({\bm \iota}^1_{5,5})=E_{5,5}^1-1\right\}\cup \left\{\left({\bf g}^1_{5,5}\right)^{{\bm \iota}^1_{5,5}}: s({\bm \iota}^1_{5,5})=E_{5,5}^1\right\},\label{g_5}\\
  &\left\{ (h_{5,b}u^b_{5,6}+h_{5,d}u^d_{5,6})\left({\bf g}^1_{5,6}\right)^{{\bm \iota}^1_{5,6}}: s({\bm \iota}^1_{5,6})=E_{5,6}^1-1\right\}\cup \left\{\left({\bf g}^1_{5,6}\right)^{{\bm \iota}^1_{5,6}}: s({\bm \iota}^1_{5,6})=E_{5,6}^1\right\},\label{g_6}\\
  &\left\{\left({\bf g}^r_{(i,j),k}\right)^{{\bm \iota}^r_{(i,j),k}}: s({\bm \iota}^r_{(i,j),k})\text{ is a constant}\right\}, \quad \forall {\bf g}^r_{(i,j),k}\notin\{{\bf g}^1_{5,k}\}_{k\in[6]}.\label{g_other}
\end{align}
\end{subequations}
Here, each $E_{5,k}^1$ denotes a specific constant value of $s({\bm \iota}^1_{5,k})$ with $k\in[6]$. Considering the grid network topology, it can be found that the polynomial structures of \eqref{g_1} \eqref{g_2} \eqref{g_3} and \eqref{g_4} are symmetric, and the polynomial structures of \eqref{g_5} and \eqref{g_6} are also symmetric. Thus, we shall only prove the linear independence of \eqref{g_1} \eqref{g_5} and \eqref{g_other}.

\subsubsection{Linear independence of the polynomial functions in \eqref{g_1}}
\begin{figure*}[t]
\begin{centering}
\includegraphics[scale=0.8]{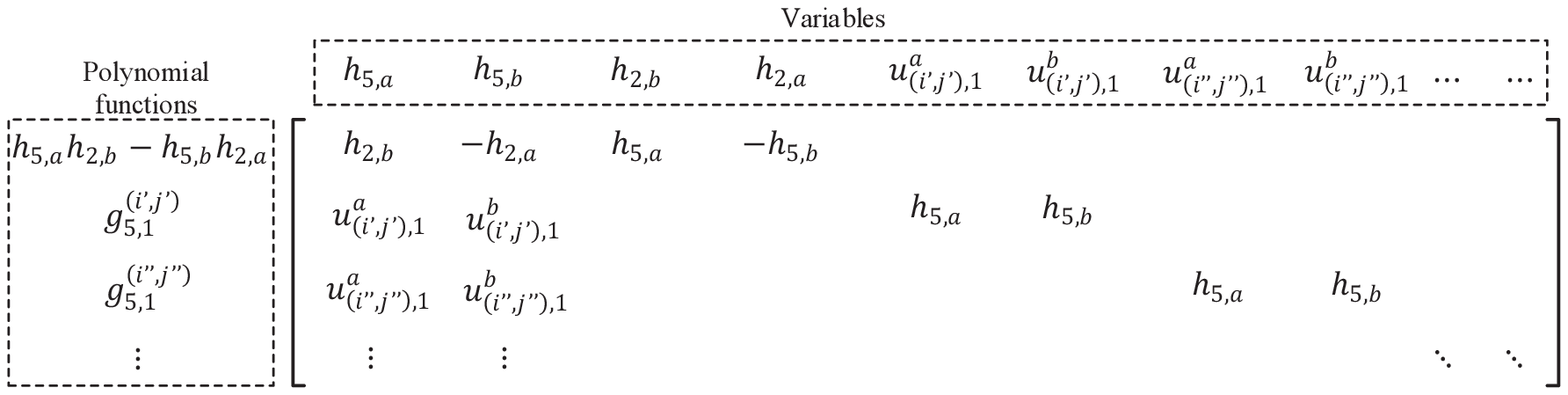}
\caption{The structure of Jacobian matrix of the polynomial functions in \eqref{polynomial_1}.} \label{jacobian_set1}
\end{centering}
\end{figure*}

Based on \cite[Corollary 1]{DoF_CoMP}, it is equivalent to prove the algebraical independence of polynomial functions in the following set
\begin{align}\label{polynomial_1}
\{(h_{5,a}h_{2,b}-h_{5,b}&h_{2,a})\}\cup\Big\{g_{5,1}^{(i',j')}=h_{5,a}u^a_{(i',j'),1}+h_{5,b}u^b_{(i',j'),1}:(i',j')\in{\mathcal U}_1\setminus\{5\}\Big\}.
\end{align}
From the construction of $g_{5,1}^{(i',j')}$, it can be seen that each of $\{g_{5,1}^{(i',j')}:(i',j')\in{\mathcal U}_1\setminus\{5\}\}$ has unique factors $u^a_{(i',j'),1}$ and  $u^b_{(i',j'),1}$. Let $\bf J$ denote the $(\frac{\sqrt{|\Psi|}}{2})\times(\sqrt{|\Psi|}+2)$ Jacobian matrix of the above polynomial functions. The structure of $\bf J$ is given in Fig. \ref{jacobian_set1}. It can be seen that $\bf J$ has full row rank with probability one. Based on \cite[Lemma 1]{DoF_CoMP} and \cite[Theorem 3, p. 135]{algebraic_independence}, the polynomial functions in \eqref{polynomial_1} are thus algebraically independent.

\subsubsection{Linear independence of the polynomial functions in \eqref{g_5}}
By using \cite[Corollary 1]{DoF_CoMP}, we shall prove the algebraical independence of polynomial functions in the following set
\begin{align}\label{polynomial_5}
\{(h_{5,a}u^a_{5,5}+h_{5,c}&u^c_{5,5})\}\cup\Big\{g_{5,5}^{(i',j')}=h_{5,a}u^a_{(i',j'),5}+h_{5,c}u^c_{(i',j'),5}:(i',j')\in{\mathcal U}_1\setminus\{5\}\Big\}.
\end{align}
The $(\frac{|\Psi|}{4})\times(\frac{|\Psi|}{2}+2)$ Jacobian matrix of the above polynomial functions has the similar structure as the one in Fig. \ref{jacobian_set1}, and thus it also has full row rank with probability one. Due to \cite[Lemma 1]{DoF_CoMP} and \cite[Theorem 3, p. 135 ]{algebraic_independence}, the polynomial functions in \eqref{polynomial_5} are algebraically independent.

\subsubsection{Linear independence of the polynomial functions in \eqref{g_other}} We divide all the sets in \eqref{g_other} into two parts, i.e, $\{({\bf g}^r_{(i,j),k})^{{\bm \iota}^r_{(i,j),k}}\}$ with $k\in[4]$, and $\{({\bf g}^r_{(i,j),k})^{{\bm \iota}^r_{(i,j),k}}\}$ with $k\in\{5,6\}$. Since the Jacobian matrix of polynomial functions $\{g_{(i,j),k}^{(i',j')}:(i',j')\in{\mathcal U}_r\setminus\{(i,j)\}\}$, with $k\in[4]$, has the similar structure as the one of $\{g_{5,1}^{(i',j')}:(i',j')\in{\mathcal U}_1\setminus\{5\}\}$, the polynomial functions $\{g_{(i,j),k}^{(i',j')}:(i',j')\in{\mathcal U}_r\setminus\{(i,j)\}\}$, with $k\in[4]$, are algebraically independent. Since the Jacobian matrix of polynomial functions $\{g_{(i,j),k}^{(i',j')}:(i',j')\in{\mathcal U}_r\setminus\{(i,j)\}\}$, with $k\in\{5,6\}$, has the similar structure as the one of $\{g_{5,5}^{(i',j')}:(i',j')\in{\mathcal U}_1\setminus\{5\}\}$, the polynomial functions $\{g_{(i,j),k}^{(i',j')}:(i',j')\in{\mathcal U}_r\setminus\{(i,j)\}\}$, with $k\in\{5,6\}$, are also algebraically independent. Thus, we can prove the linear independence of the polynomial functions within each set in \eqref{g_other}.

\section{Proof of Theorem \ref{DoF_upper_theorem}}\label{proof DoF upper theorem}
Define the following subfile sets
\begin{subequations}
   \begin{align}
  {\mathcal W}_{\mathcal R,\Phi}&=\left\{W_{r_{(i,j)},{{\mathcal A}_k}}: (i,j) \in {\mathcal R}, {{\mathcal A}_k} \in \Theta \right\}, \\
  {\mathcal W}_{\bar{\mathcal R},\mathcal T}&=\left\{W_{r_{(i,j)},{{\mathcal A}_k}}: (i,j) \in \bar{\mathcal R}, {{\mathcal A}_k} \subseteq {\mathcal T}, {{\mathcal A}_k} \in \Theta \right\},
\end{align}
\end{subequations}
and define $\bar{\mathcal W}$ as the set of the remaining subfiles to be sent in the network. For a user set ${\mathcal R}$, we use ${\bf y}_{\mathcal R}$ to denote the vector of the signals received by the users in ${\mathcal R}$, with a similar notation used for the transmitted signals and noise. In the following, we show that given ${\bf y}_{\mathcal R}$, the subfile sets ${\mathcal W}_{\mathcal R,\Phi}$ and ${\mathcal W}_{\bar{\mathcal R},{\mathcal T}}$ can be recovered with $\bar{\mathcal W}$ as side information, i.e.
 \begin{equation}\label{boundentropy1}
  H({\mathcal W}_{\mathcal R,\Phi},{\mathcal W}_{\bar{{\mathcal R}}, {\mathcal T}}|{\bf y}_{\mathcal R},\bar{\mathcal W})\leq \epsilon
\end{equation}
Based on the converse assumption that each user can obtain the desired file from its received signal, ${\mathcal W}_{\mathcal R,\Phi}$ can be decoded from ${\bf y}_{\mathcal R}$ almost surely. The signals ${\bf y}_{\mathcal R}$ and ${\bf y}_{\bar{\mathcal R}}$ can be expressed as
\begin{subequations}
\begin{align}
  {\bf y}_{\mathcal R}={\bf H}_{\mathcal R,\mathcal T}{\bf x}_{\mathcal T}+{\bf H}_{\mathcal R,\Phi \setminus \mathcal T}{\bf x}_{\Phi \setminus \mathcal T}+{\bf n}_{\mathcal R},\\
  {\bf y}_{\bar{\mathcal R}}={\bf H}_{\bar{\mathcal R},\mathcal T}{\bf x}_{\mathcal T}+{\bf H}_{\bar{\mathcal R},\Phi \setminus \mathcal T}{\bf x}_{\Phi \setminus \mathcal T}+{\bf n}_{\bar{\mathcal R}},
\end{align}
\end{subequations}
 respectively. From ${\mathcal W}_{\mathcal R,\Phi}$ and $\bar{\mathcal W}$, we can obtain those requested subfiles which are cached at BSs in set $\Phi\setminus{\mathcal T}$ and further reconstruct the transmitted signal ${\bf x}_{\Phi \setminus \mathcal T}$. Utilizing ${\bf x}_{\Phi \setminus \mathcal T}$ and ${\bf y}_{\mathcal R}$, we can obtain a degraded version of ${\bf y}_{\bar{\mathcal R}}$ as
\begin{align}
  \tilde{{\bf y}}_{\bar{\mathcal R}}&={\bf H}_{\bar{\mathcal R},\mathcal T}{\bf H}^\dag_{\mathcal R,\mathcal T}({\bf y}_{\mathcal R}-{\bf H}_{\mathcal R,\Phi \setminus \mathcal T}{\bf x}_{\Phi \setminus \mathcal T})+{\bf H}_{\bar{\mathcal R},\Phi \setminus \mathcal T}{\bf x}_{\Phi \setminus \mathcal T}\nonumber\\
  &={\bf H}_{\bar{\mathcal R},\mathcal T}{\bf x}_{\mathcal T}+{\bf H}_{\bar{\mathcal R},\Phi \setminus \mathcal T}{\bf x}_{\Phi \setminus \mathcal T}+{\bf H}_{\bar{\mathcal R},\mathcal T}{\bf H}^\dag_{\mathcal R,\mathcal T}{\bf n}_{\mathcal R}.
\end{align}
Since $\tilde{{\bf y}}_{\bar{\mathcal R}}$ and ${\bf y}_{\bar{\mathcal R}}$ only differ in noise, the subfile set ${\mathcal W}_{\bar{\mathcal R},{\mathcal T}}$ can be recovered from $\tilde{{\bf y}}_{\bar{\mathcal R}}$ almost surely.

By using Fano's inequality, we have
\begin{subequations}\label{bound_entropy}
\begin{align}
   H({\mathcal W}_{\mathcal R,\Phi},{\mathcal W}_{\bar{{\mathcal R}}, {\mathcal T}}) &=\left(\sum\limits_{(i,j)\in {\mathcal R}}\sum\limits_{{\mathcal A}_k \in \Theta}R_{r_{(i,j)},{{\mathcal A}_k}}+\sum\limits_{(i,j)\in \bar{\mathcal R}}\sum\limits_{{\mathcal A}_k \subseteq {\mathcal T}, {{\mathcal A}_k} \in \Theta}R_{r_{(i,j)},{{\mathcal A}_k}}\right)\cdot T\\
   &= H({\mathcal W}_{\mathcal R,\Phi},{\mathcal W}_{\bar{{\mathcal R}}, {\mathcal T}}|\bar{\mathcal W}) \label{equalityofmessage}\\
   &\leq I({\mathcal W}_{\mathcal R,\Phi},{\mathcal W}_{\bar{{\mathcal R}}, {\mathcal T}};{\bf y}_{\mathcal R}|\bar{\mathcal W}) +H({\mathcal W}_{\mathcal R,\Phi},{\mathcal W}_{\bar{{\mathcal R}}, {\mathcal T}}|{\bf y}_{\mathcal R},\bar{\mathcal W}) \\
   &\leq h({\bf y}_{\mathcal R}|\bar{\mathcal W})-h({\bf y}_{\mathcal R}|{\mathcal W}_{\mathcal R,\Phi},{\mathcal W}_{\bar{{\mathcal R}}, {\mathcal T}},\bar{\mathcal W})+\epsilon \\
   &\leq h({\bf y}_{\mathcal R})+\epsilon \label{minus}\\
   &\leq T|\mathcal R|\log P+\epsilon \label{boundentropy2}
\end{align}
\end{subequations}
where \eqref{equalityofmessage} comes from the fact that $\bar{\mathcal W}$ is independent from ${\mathcal W}_{\mathcal R,\Phi}$ and ${\mathcal W}_{\bar{{\mathcal R}}, {\mathcal T}}$; \eqref{boundentropy2} is due to \cite[Lemma 5]{long_journal}. By rearranging \eqref{bound_entropy} and taking $T\rightarrow \infty$ and $\epsilon\rightarrow 0$, we have
\begin{align}
      \sum\limits_{(i,j)\in {\mathcal R}}\sum\limits_{{{\mathcal A}_k} \in \Theta}d_{(i,j),{{\mathcal A}_k}}+\sum\limits_{(i,j)\in \bar{\mathcal R}}\sum\limits_{{{\mathcal A}_k} \subseteq {\mathcal T}, {{\mathcal A}_k} \in \Theta}d_{(i,j),{{\mathcal A}_k}}\leq |\mathcal R|.
\end{align}
For all the pairs $(\mathcal R,\mathcal T)$ in $\mathcal S$, the above condition must be satisfied, which completes the proof.

\section{Proof of Lemma \ref{bs_t}}\label{proof_bs_t}
 We need to prove $\lambda(\mathcal R,\mathcal T)\leq\lambda(\mathcal R^*,\mathcal T^*)$, where $\mathcal T^*$ is given in Lemma \ref{bs_t} and $\mathcal R^*$ is the corresponding user set for $\mathcal T^*$. First, we consider $\mathcal T$ not entirely including any ${\mathcal B}_k$. According to the adopted cache placement scheme, each BS cache group ${\mathcal A}_k$ contains at least one ${\mathcal B}_k$. Thus, we have
 \begin{align}
   \lambda(\mathcal R,\mathcal T)=\frac{|\bar{\mathcal R}|}{{d|\mathcal R|}}\sum\limits_{{{\mathcal A}_k} \subseteq {\mathcal T}, {{\mathcal A}_k} \in \Theta}d_{(i,j),{{\mathcal A}_k}}=\frac{1}{{d|\mathcal R|}}\cdot0=0.
 \end{align}
  It is obvious that $\lambda(\mathcal R,\mathcal T)\leq\lambda(\mathcal R^*,\mathcal T^*)$. Then, we consider $\mathcal T$ including $\{{\mathcal B}_k\}_{k\in{\mathcal Q}}$ and some other BSs, where $\mathcal Q\subseteq[4]$, e.g., $\mathcal T={\mathcal B}_1\cup\{(0,2)\}$. We shall prove $\lambda(\mathcal R,\mathcal T)\leq\lambda(\mathcal R^*,\mathcal T^*)$ with $\mathcal T^*=\{{\mathcal B}_k\}_{k\in{\mathcal Q}}$. Since in the adopted caching scheme each BS cache group is a union of some ${\mathcal B}_k$'s, we have $\sum\limits_{{{\mathcal A}_k} \subset {\mathcal T}, {{\mathcal A}_k} \in \Theta}d_{(i,j),{{\mathcal A}_k}}=\sum\limits_{{{\mathcal A}_k} \subseteq {\mathcal T^*}, {{\mathcal A}_k} \in \Theta}d_{(i,j),{{\mathcal A}_k}}$. Noticing that both $\mathcal T$ and $\mathcal T^*$ are connected to all users $\Psi$, we have $|\mathcal R|+|\bar{\mathcal R}|=|\mathcal R^*|+|\bar{\mathcal R^*}|$. Since $|\bar{\mathcal R}^*|\geq|\bar{\mathcal R}|$, we can obtain
  \begin{align}
    \lambda(\mathcal R,\mathcal T)&=\frac{|\bar{\mathcal R}|}{{d|\mathcal R|}}\sum\limits_{{{\mathcal A}_k} \subset {\mathcal T}, {{\mathcal A}_k} \in \Theta}d_{(i,j),{{\mathcal A}_k}}\nonumber \\
    &\leq\frac{|\bar{\mathcal R^*}|}{{d|\mathcal R^*|}}\sum\limits_{{{\mathcal A}_k} \subseteq {\mathcal T^*}, {{\mathcal A}_k} \in \Theta}d_{(i,j),{{\mathcal A}_k}}\nonumber \\
    &=\gamma(\mathcal R^*,\mathcal T^*).
  \end{align}

\section{Proof of Theorem \ref{theorem3}}\label{proof_theorem3}

\subsection{$\mu\in\left[\frac{1}{4},\frac{1}{2}\right)$}
In this case, memory sharing is used between cache sizes $\mu=\frac{1}{4}$ and $\mu=\frac{1}{2}$, i.e., $\mu=\frac{\gamma}{4}+\frac{1-\gamma}{2}$, where $\gamma\in(0,1]$. It means that $\gamma F$ bits of each file are cached in the manner of point $\mu=\frac{1}{4}$, and the remaining $(1-\gamma)F$ bits of each file are cached in the manner of point $\mu=\frac{1}{2}$. Denote $d_{\frac{1}{4}}$ and $d_{\frac{1}{2}}$ as the per-user DoF for subfiles cached in the manner of point $\mu=\frac{1}{4}$ and point $\mu=\frac{1}{2}$, respectively. Thus, we have $d_{\frac{1}{4}}(1-\gamma)=d_{\frac{1}{2}}\gamma$ and $d_{\frac{1}{4}}+d_{\frac{1}{2}}=d$. By taking ${\mathcal T}^*$ in Lemma \ref{bs_t} and the corresponding ${\mathcal R}^*$ into \eqref{inequality_temp}, we can obtain
 \begin{align}
 \begin{cases}
    \frac{7}{4}d_{\frac{1}{4}}+d_{\frac{1}{2}}\leq 1, & \text{for}\ {\mathcal T}^*\in\{{\mathcal B}_1,{\mathcal B}_2,{\mathcal B}_3,{\mathcal B}_4\},\\
   \frac{3}{2}d_{\frac{1}{4}}+\frac{7}{6}d_{\frac{1}{2}}\leq 1, & \text{for}\ {\mathcal T}^*\in\{{\mathcal B}_1\cup{\mathcal B}_2,{\mathcal B}_1\cup{\mathcal B}_3,{\mathcal B}_1\cup{\mathcal B}_4,{\mathcal B}_2\cup{\mathcal B}_3,{\mathcal B}_2\cup{\mathcal B}_4,{\mathcal B}_3\cup{\mathcal B}_4\}, \\
   \frac{5}{4}d_{\frac{1}{4}}+\frac{7}{6}d_{\frac{1}{2}}\leq 1, & \text{for}\ {\mathcal T}^*\in\{\Phi\backslash{\mathcal B}_1,\Phi\backslash{\mathcal B}_2,\Phi\backslash{\mathcal B}_3,\Phi\backslash{\mathcal B}_4\},\\
   d_{\frac{1}{4}}+d_{\frac{1}{2}}\leq 1, & \text{for}\ {\mathcal T}^*=\Phi.\\
 \end{cases}
 \end{align}
By mathematical deduction, we can obtain $d\leq\min\left\{\frac{2}{5-6\mu},\frac{6}{11-8\mu}\right\}$.

\subsection{$\mu\in\left[\frac{1}{2},1\right]$}
We use memory sharing between cache sizes $\mu=\frac{1}{2}$ and $1$ in this region, i.e., $\mu=\frac{\gamma}{2}+(1-\gamma)$, where $\gamma\in[0,1]$. Denote $d_{1}$ as the per-user DoF for the subfiles cached in the manner of point $\mu=1$. We have $d_{\frac{1}{2}}(1-\gamma)=d_{1}\gamma$ and $d_{\frac{1}{2}}+d_{1}=d$. By taking ${\mathcal T}^*$ in Lemma \ref{bs_t} and the corresponding ${\mathcal R}^*$ into \eqref{inequality_temp}, we can obtain
 \begin{align}
 \begin{cases}
   d_{\frac{1}{2}}+d_{1}\leq 1, & \text{for}\ {\mathcal T}^*\in\{{\mathcal B}_1,{\mathcal B}_2,{\mathcal B}_3,{\mathcal B}_4\},\\
   \frac{7}{6}d_{\frac{1}{2}}+d_{1}\leq 1, & \text{for}\ {\mathcal T}^*\in\{{\mathcal B}_1\cup{\mathcal B}_2,{\mathcal B}_1\cup{\mathcal B}_3,{\mathcal B}_1\cup{\mathcal B}_4,{\mathcal B}_2\cup{\mathcal B}_3,{\mathcal B}_2\cup{\mathcal B}_4,{\mathcal B}_3\cup{\mathcal B}_4\}, \\
    \frac{7}{6}d_{\frac{1}{2}}+d_{1}\leq 1, & \text{for}\ {\mathcal T}^*\in\{\Phi\backslash{\mathcal B}_1,\Phi\backslash{\mathcal B}_2,\Phi\backslash{\mathcal B}_3,\Phi\backslash{\mathcal B}_4\}\\
   d_{\frac{1}{2}}+d_{1}\leq 1, & \text{for}\ {\mathcal T}^*=\Phi.\\
 \end{cases}
 \end{align}
By mathematical deduction, we can obtain $d\leq\frac{3}{4-\mu}$.

\end{appendices}
\bibliographystyle{IEEEtran}
\bibliography{IEEEabrv,reference}

\end{document}